\documentclass[12pt,twoside,a4paper]{article}
\usepackage{amsmath,amssymb,latexsym,theorem,natbib,epsfig,color,subfigure,footmisc,bbm}
\usepackage{multirow,graphicx,array,rotating,url,multicol}  
\usepackage{epstopdf,afterpage,wasysym,hyperref}
\usepackage{verbatim}
\usepackage[normalem]{ulem}
\def\crps{\mathop{\hbox{\rm CRPS}}}
\def\crpss{\mathop{\hbox{\rm CRPSS}}}
\def\mae{\mathop{\hbox{\rm MAE}}}
\def\maes{\mathop{\hbox{\rm MAES}}}
\def\ri{\mathop{\hbox{\rm RI}}}
\def\qs{{\mathrm {QS}}}
\def\rel{{\mathrm {REL}}}
\def\res{{\mathrm {RES}}}
\def\unc{{\mathrm {UNC}}}

\title{Post-processing of ensemble photovoltaic power forecasts with distributional and quantile regression methods}

\author{{Martin J\'anos Mayer}$^{1}$, {\'Agnes Baran}$^{2}$, {Sebastian Lerch}$^{3,4}$, \\ {Nina Horat}$^{5}$, {Dazhi Yang}$^{6}$ and {S\'andor Baran}$^{2}$ \vspace*{0.5cm}\\
{\small $^1$Department of Energy Engineering, Faculty of Mechanical Engineering, Budapest}\\ {\small University of Technology and Economics, Hungary}\\ 
{\small $^2$Faculty of Informatics, University of Debrecen, Hungary} \\
{\small $^3$Department of Mathematics and Computer Science, Marburg University,
  Germany} \\
{\small $^4$Heidelberg Institute for Theoretical Studies, Germany} \\
{\small $^5$Institute of Statistics, Karlsruhe Institute of Technology, Germany} \\
{\small $^6$School of Electrical Engineering and Automation, Harbin Institute of Technology,}\\ {\small Harbin, Heilongjiang, China}
}

\date{}

\begin{document}
\maketitle

\begin{abstract}
Accurate and reliable forecasting of photovoltaic (PV) power generation is crucial for grid operations, electricity markets, and energy planning, as solar systems now contribute a significant share of the electricity supply in many countries. PV power forecasts are often generated by converting forecasts of relevant weather variables to  power predictions via a model chain. The use of ensemble simulations from numerical weather prediction models results in probabilistic PV forecasts in the form of a forecast ensemble. However, weather forecasts often exhibit systematic errors that propagate through the model chain, leading to biased and/or uncalibrated PV power predictions. These deficiencies can be mitigated by statistical post-processing.  Using PV production data and corresponding short-term PV power ensemble forecasts at seven utility-scale PV plants in Hungary, we systematically evaluate and compare seven state-of-the-art methods for post-processing PV power forecasts. These include both parametric and non-parametric techniques, as well as statistical and machine learning-based approaches. Our results show that compared to the raw PV power ensemble, any form of statistical post-processing significantly improves the predictive performance. Non-parametric methods outperform parametric models, with advanced nonlinear quantile regression models showing the best results. Furthermore, machine learning-based approaches surpass their traditional statistical counterparts.

\end{abstract}

\section{Introduction}
\label{sec1}
The global shift toward low-carbon energy systems has brought renewable energy sources to the forefront of electricity generation \citep{hong2020energy,y22ras}. Among these, solar photovoltaic (PV) power has emerged as a key pillar of sustainable energy strategies due to its scalability, declining costs, and widespread deployment, with PV systems now contributing a significant share of electricity supply in many countries \citep{ywx22cons}. However, the inherently intermittent nature of solar energy presents substantial challenges, in particular for maintaining the stability and efficiency of power systems. Accurate and reliable forecasting of PV electricity generation thus plays a critical role in grid operations, electricity markets, and energy planning. In recent years, a growing emphasis has been placed on probabilistic forecasting approaches \citep{ym21pp,gls23}. These methods move beyond single-valued point predictions by providing comprehensive information in the form of prediction intervals, quantiles, of full probability distributions, and thus enabling the quantification of forecast uncertainty \citep{ldp19,Gneiting2023}.

PV power forecasts are often generated following a three-stage framework, where weather forecasts of global horizontal irradiance (GHI) and other variables from a weather forecasting model are converted to PV power forecasts via a model chain \citep{Roberts2017, my22mce, yxm24tut1}. The weather forecasts, which serve as key inputs, are nowadays usually based on numerical weather prediction (NWP) models, which describe physical processes via systems of differential equations. Ensemble simulations from NWP systems with varying initial conditions or model physics enable the quantification of forecast uncertainties and serve as a straightforward baseline method for generating probabilistic PV power forecasts by applying the PV model chain conversion individually to all ensemble members.

Although this approach allows for propagating forecast uncertainty through the model chain, there is broad evidence that NWP ensemble predictions often show systematic errors \citep{b18b}. In particular, they are often subject to systematic biases and fail to reliably quantify forecast uncertainty for many variables. 
Therefore, they require correction via so-called post-processing methods. These are statistical or machine learning (ML)-based distributional regression models, which yield probabilistic forecasts in the form of probability distributions, quantiles, or corrected ensemble predictions, see \citet{vbd21} for an overview of recent developments. A particular focus of recent research in post-processing has been on the development of modern ML methods such as random forests \citep{tmzn16qrf} or neural networks \citep[NNs;][]{rl18}. By allowing for incorporating multiple meteorological variables as inputs and flexibly modeling nonlinear relationships, they have been found to yield substantial improvements in predictive performance over classical approaches based on statistical methods in various applications, see, e.g., \citet{haupt_etal_2021_implementing,vbd21,sl22,demaeyer_etal_2023_euppbench} for overviews and comparisons.

Over the last years, post-processing methods have also been applied to solar energy prediction, in particular for post-processing NWP forecasts of solar irradiance \citep{Bakker2019,lsb20,sealb21,ym21pp,bb24,sy24ncqrnn}.
The review by \cite{ym21pp} categorized post-processing methods into four groups based on the deterministic or probabilistic nature of the inputs and outputs, namely D2D, D2P, P2D, and P2P. While the paper noted that P2P post-processing was, by that time, the least developed category among the four, and although there has been notable progress in the last four years, P2P post-processing is still less established compared to the deterministic post-processing methods.

In the context of model chain approaches to PV power prediction, P2P post-processing can be applied at different stages, following one of four possible strategies: Propagating the raw, unprocessed ensemble weather predictions through the model chain without applying any post-processing; applying post-processing only to the weather inputs before the conversion to PV power; applying post-processing only to the PV power forecasts obtained through the model chain conversion; or applying post-processing both before and after the conversion. \citet{hkl25} compared these strategies using statistical and ML-based post-processing methods based on a benchmark dataset \citep{WangEtAl2022} and found that post-processing the PV power predictions is the most promising strategy, which is in line with results from related research on deterministic solar energy prediction \citep{Theocharides2020,my24opt} and on probabilistic wind power forecasting \citep{PhippsEtAl2022}. 
Further, \citet{hkl25} noted that ML-based post-processing methods outperform their statistical counterparts for solar energy forecasting, albeit by a relatively small margin. 

Our overarching aim is to systematically evaluate and compare statistical and ML-based P2P post-processing methods for the calibration and conditional bias correction of PV power forecasts. 
Following the findings of \citet{hkl25}, we focus on comparing post-processing methods for ensemble forecasts of PV power obtained as the output of the model chain conversion step when using raw NWP ensemble predictions as inputs. Overall, we compare seven methods with a particular focus on parametric distributional regression approaches, which assume a parametric family of probability distributions for the target variable, and non-parametric quantile regression methods, which yield a set of quantiles as their output.

The investigated parametric distributional regression methods include the ensemble model output statistics \citep[EMOS;][]{grwg05} approach, which is also referred to as non-homogeneous regression and was originally proposed with the assumption of a Gaussian forecast distribution, but has been extended towards solar energy forecasting \citep{sealb21,hkl25}. \citet{mmz17} proposed a gradient boost-based extension of EMOS, which enables the incorporation of additional predictor variables and which we will refer to as EMOS-B or boosted EMOS. A neural network-based distributional regression approach to post-processing was proposed by \citet{rl18} and will be referred to as distributional regression network (DRN). 

A key drawback of parametric distributional regression approaches is the need to choose a suitable parametric family for the conditional distribution of the variable of interest, given the ensemble predictions. Non-parametric methods circumvent this disadvantage, with quantile-regression based methods constituting the most popular approach. We here compare standard linear quantile regression to quantile regression networks \citep[QRNNs;][]{t00qrnn}, where neural networks are used to learn non-linear mappings from the input predictors to target quantiles. 
We further consider Bernstein quantile network \citep[BQNs;][]{b20bqn}, which model the quantile function as a weighted mixture of Bernstein polynomials, as well as the recent non-crossing quantile regression neural network (NCQRNN) approach proposed by \citet{sy24ncqrnn}, which modifies the QRNN architecture to avoid quantile crossing.

Our comparisons are based on a five-year dataset of PV production at seven utility-scale PV plants in Hungary and corresponding ensemble weather forecasts, and thus notable extend the scale of the comparisons conducted in \citet{hkl25} both in terms of the amount of data, but also the breadth of post-processing methods.
Specifically, the comparison of parametric and non-parametric approaches allows for assessing the challenges of choosing a suitable parametric family for PV power production, while considering both classical statistical as well as modern ML-based methods enables insights into the benefits of the potential to flexibly learn non-linear relationships via NNs. 

The remainder of this article is organized as follows. 
Section \ref{sec2} provides a comprehensive description of the PV power plant and weather forecast data used in this study, as well as the specifics of the model chain which is used for the conversion from weather to PV power forecasts. In Section \ref{sec3},  we introduce the post-processing approaches and forecast evaluation methods. Section \ref{sec4} presents the main results, and Section \ref{sec5} concludes with a discussion. Additional results can be found in the Appendix.

\begin{figure}[t]
\begin{center}
\epsfig{file=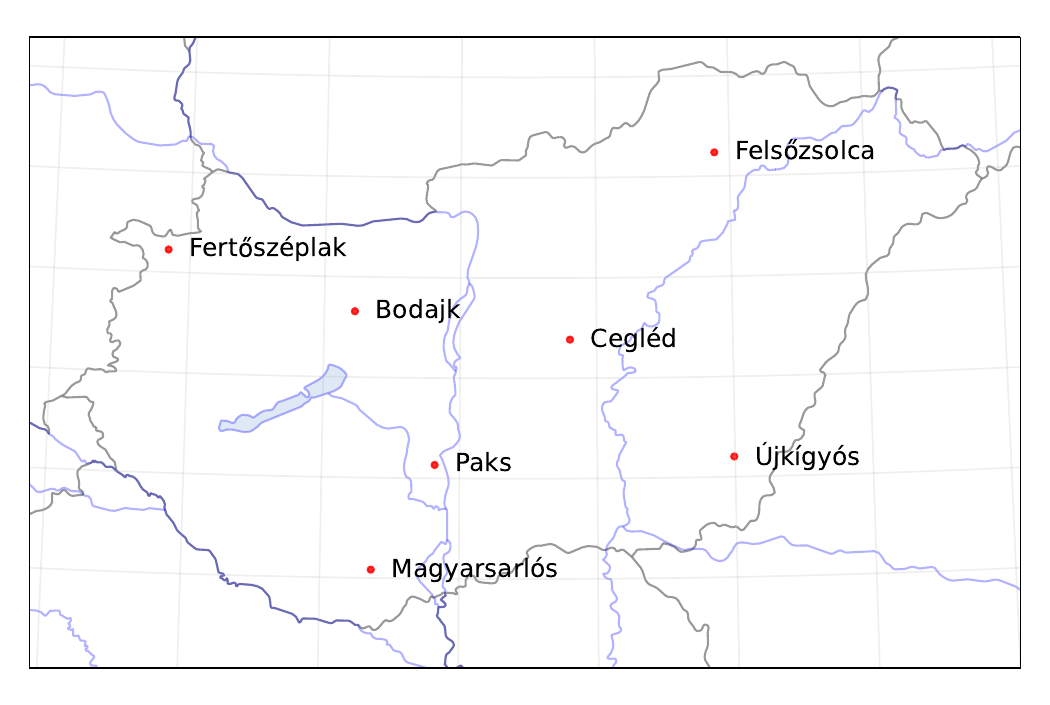, width=.5\textwidth}
\end{center}
\caption{Locations of the seven utility-scale PV plants considered in this study}
\label{fig:locations}
\end{figure}

\section{Photovoltaic power production and forecast data}
\label{sec2}
The post-processing models are tested for the operational day-ahead power forecasting at seven PV plants in Hungary. The input is a 51-member PV power forecast ensemble, created by converting all members of an ensemble NWP weather forecast into PV power by a physical PV model chain. The description of the PV plant data, the ensemble NWP, and the model chain are provided in the following subsections.

\subsection{Photovoltaic power plant data}
\label{subs2.1}
The PV power forecasting is performed for seven ground-mounted utility-scale PV plants in Hungary. The locations of the PV plants are shown in the map of Hungary in Figure \ref{fig:locations}, and their geographical coordinates and main design parameters are summarized in Table \ref{tab:locations}. The measured power output data of the PV plants are available for the five full calendar years from 2019 to 2023 with a temporal resolution of 15 min, which fits the operational requirements for scheduling PV plants in Hungary. Only daytime data are considered in this study, selected by a zenith angle $\Theta_z <90^{\circ}$ filter, and the daytime data samples with 0 power output are removed from the dataset as they indicate the malfunction or maintenance of the PV plants. The number of valid daytime data samples that are used in the analysis is presented for each year and PV plant in Table \ref{tab:locations}. Furthermore, to handle the capacity differences of the investigated power plants, both the measured output PV power at a given location and the corresponding PV power forecast are normalized by the nominal AC power of the plant at hand provided in Table \ref{tab:locations}.

\begin{table}[t]  
  \begin{center}{\footnotesize
    \begin{tabular}{l|cc|cc|cc|ccccc}
       \hline
&\multicolumn{2}{|c|}{Geographical}&\multicolumn{2}{c|}{Module}& \multicolumn{2}{c|}{Nominal}&\multicolumn{5}{c}{Number of valid 15-min} \\
Name &\multicolumn{2}{|c|}{location}&\multicolumn{2}{c|}{orientation}& \multicolumn{2}{c|}{power (kW)}&\multicolumn{5}{c}{daytime data samples} \\\cline{2-12} 
&Lat.&Lon.&Tilt&Azim.&DC&AC&2019&2020&2021&2022&2023 \\ \hline
Bodajk&$47.33^\circ$&$18.22^\circ$&$35^\circ$&$180^\circ$&590&498&17530&17547&17579&	17541&17533\\
Cegl\'ed&$47.19^\circ$&$19.80^\circ$&$35^\circ$&$180^\circ$&590&498&17067&17015&17158&	17090&17058\\
Fels\H{o}zsolca&$48.12^\circ$&$20.89^\circ$&$35^\circ$&$180^\circ$&20038&18640&17331&	17337&17432&17341&16934 \\
Fert\H{o}sz\'eplak&$47.61^\circ$&$16.84^\circ$&$35^\circ$&$180^\circ$&590&498&17533&	17583&17572&17506&17505\\
Magyarsarl\'os&$46.04^\circ$&$18.37^\circ$&$25^\circ$&$160^\circ$&601&502&17616&17561&	17568&17550&17510\\
Paks&$46.57^\circ$&$18.82^\circ$&$35^\circ$&$180^\circ$&20680&19160&17213&17188&17273&	17052&17163\\
\'Ujk\'{\i}gy\'os&$46.60^\circ$&$20.99^\circ$&$35^\circ$&$180^\circ$&590&498&17095&16924& 17045&17027&17066\\\hline
       \end{tabular}
       }
    \end{center}
    \caption{Description of the PV plants considered in this study.}
    \label{tab:locations}
\end{table}

\subsection{Ensemble numerical weather predictions}
\label{subs2.2}
The weather input data for the PV power forecasts are retrieved from the ensemble (ENS) NWP product of the Integrated Forecasting System (IFS) of the European Centre for Medium-Range Weather Forecasts (ECMWF). The ECMWF ENS is an ensemble of 51 members, including a control forecast and 50 perturbed members. The initial conditions of the NWP, reflecting the current state of the Earth system, are determined by a four-dimensional variational data assimilation (4D-Var) method combining the observations and the latest short-range weather forecasts. The control forecast is created from the best available data using the unperturbed models, whereas the perturbed members are calculated from slightly changed initial conditions with slightly modified model parameterizations.  
The forecasts of all 51 members for the 24-48-h time horizon are taken from the 00 UTC model run, which is the latest model run that fits the operational requirements of day-ahead forecasting in Hungary. In June 2025, the forecasts of 00 UTC run are made available at 06:44 UTC for the next day\footnote{https://confluence.ecmwf.int/display/DAC/Dissemination+schedule}, leaving enough time to prepare the PV power forecasts before the gate closure time of the day-ahead market (DAM) of the Hungarian power exchange (HUPX) at 12:00 CET/CEST.

The forecast weather variables include the GHI (ssrd, surface solar radiation downwards), the ambient temperature at 2 meters (t2m), and the wind speed at 10 meters (norm of the v10 and u10 components). The ECMWF ENS had an 18 km spatial resolution until the Cycle 48r1 model upgrade\footnote{https://www.ecmwf.int/en/about/media-centre/news/2023/model-upgrade-increases-skill-and-unifies-medium-range-resolutions} on 27 June 2023 and 9 km afterwards, and the data from the nearest grid point to the plant location are used for each PV plant. The forecasts are available with a 1-h temporal resolution, which was downscaled to a 15-min resolution to fit the requirements of the Hungarian Transmission System Operator for day-ahead scheduling. The ambient temperature and wind speed are downsampled by linear interpolation, whereas clear-sky interpolation is used for the GHI to better retain the natural daily trend of the solar radiation. Thereby, the linear interpolation is performed on the clear sky index, calculated as the ratio of the GHI and its clear-sky counterpart, which is obtained from the McClear service \citep{Lefvre2013McClear}.

\subsection{Physical photovoltaic model chain}
\label{subs2.3}
The weather forecasts of all ensemble members are converted to PV power forecasts using a physical model chain of the PV plants. The PV model chain is a series of physical models, each describing an individual phenomenon \citep{mg20opt}. The conversion of weather data to PV power is also called solar power curve modeling; more details on the variety of the existing methods can be found in a recent tutorial review \citep{yxm24tut1}. Model chains can be constructed with different accuracy and complexity depending on the number of steps and the component models selected in each step \citep{mg21mc}. In this study, we opt for a detailed model chain in order to account for most of the nonlinearities of the energy conversion to provide the most accurate inputs for the post-processing.

\begin{figure}[t]
\begin{center}
\epsfig{file=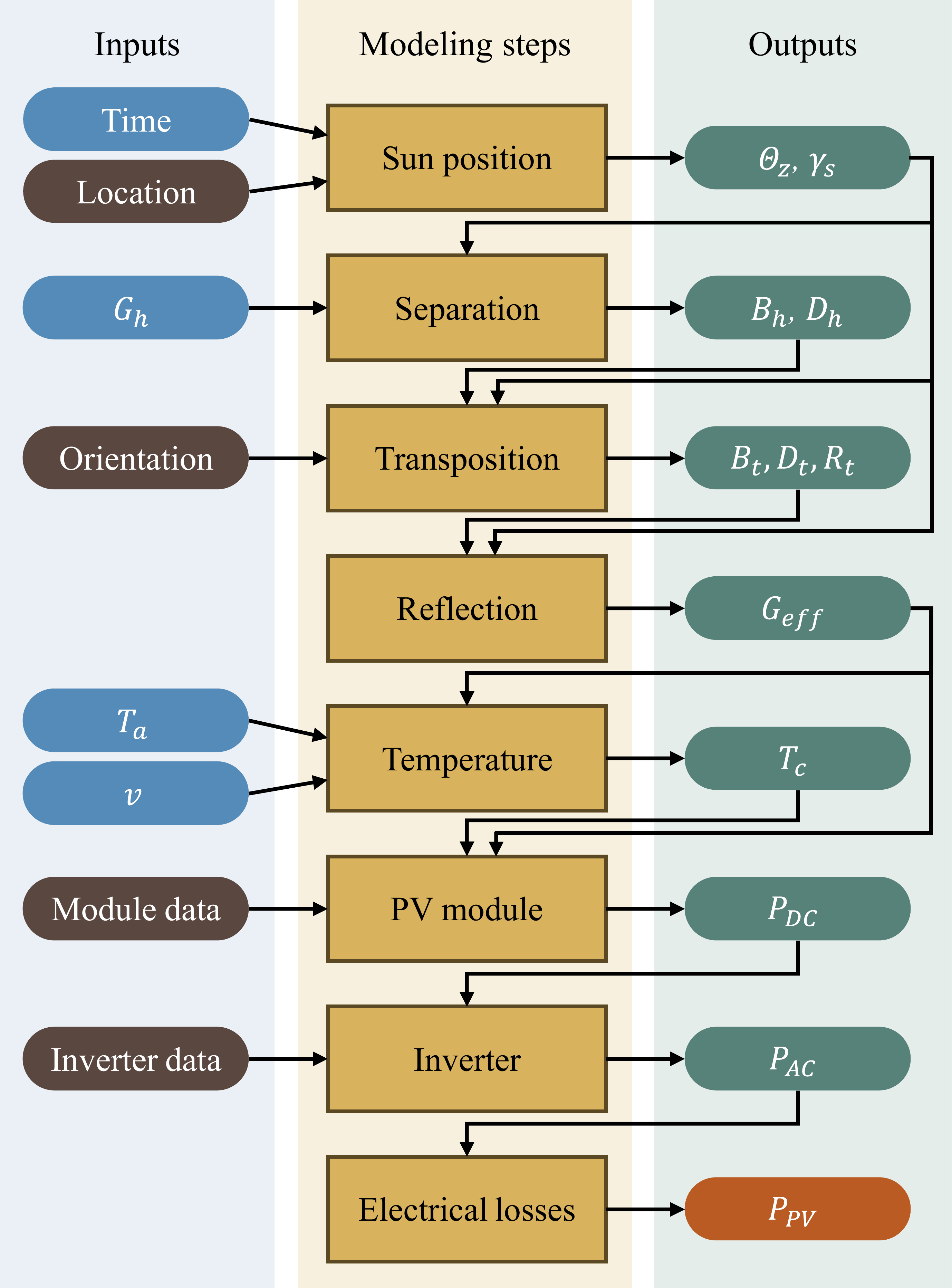, width=.5\textwidth}
\end{center}
\caption{Schematic of the physical PV model chain used for converting weather data to PV power}
\label{fig:modelchain}
\end{figure}

A schematic of the model chain implemented in this study, including the considered modeling steps along with their main inputs and outputs, is shown in Figure \ref{fig:modelchain}. A summary of the models and parameters used in the model chain is 
provided below, while the detailed description of the models can be found in the original publications of the models. The first step of the model chain is the calculation of the solar zenith and solar azimuth angles using the solar positioning algorithm of \citet{ra04spa}. It is followed by separation modeling, where the GHI is decomposed into its beam and diffuse horizontal components using the temporal-resolution cascade \textsc{Yang} model \citep{Yang2021septrc}, which emerged as one of the best separation models in a recent worldwide review \citep{Yang22seprew}. The model is used with the parameters proposed for cluster 5 in \citet{Yang2024sepclust}, since all locations at hand fall into this one out of the five clusters identified based on cloud cover frequency, aerosol optical depth, and surface albedo climatology.

The next step is to transpose the horizontal irradiance components to the tilted plane of PV arrays. The \textsc{Perez} model \citep{Perez1990trans} is selected for this task, which has been widely regarded as the most accurate transposition model for the last more than three decades \citep{Yang16transrew}. The reflection and absorption losses of the PV module cover are accounted for using the angular loss factors proposed by \citet{MartinRuiz2001}. In addition to the angular loss factor of the beam irradiance that depends on its incidence angle, this model also provides formulae for the sky-diffuse and ground-reflected components. The PV plants at hand feature a mounting structure layout of multiple parallel rows; therefore, the shading losses are estimated by assuming a 2D shading geometry, considering the nonlinear response of the PV power output to the shaded area, as described in \citet{mg20opt}. The masking of the diffuse irradiance by the adjacent rows is also taken into account by a reduced sky view factor \citep{m22impact}.

The temperature of the solar cells is calculated using the \textsc{Mattei} model, which includes both the heating of the modules due to the absorbed radiation and the effect of wind speed on the heat transfer coefficient \citep{Mattei2006}. The power output of the PV modules is calculated using the 5-parameter single-diode equivalent circuits of the modules, as described by \citet{desoto06}. First, the parameter values at nominal conditions are determined based on the datasheet of the PV modules, then they are corrected for the actual irradiance and cell temperature and used to plot the whole I-V characteristic curve of the modules for each timestep. The current, voltage, and power output of the PV modules are obtained from the maximum power point of the characteristic curve. The degradation of the module is covered by an initial 2\,\% light-induced and a 0.5\,\%/a annual loss factor.

The input power and voltage of the inverter are calculated considering the string layout of the modules and the DC cable losses. The inverter efficiency is estimated as a function of both the input power and voltage using the \textsc{Driesse} model \citep{driesse08} with parameters fitted to the efficiency values disclosed in the datasheet of the inverters. The clipping losses are also accounted for by maximizing the power output at the nominal AC power of the inverter. Finally, the power is summarized for all inverters, and electric losses on the AC cables, transformers, and other components are deducted to find the power fed into the grid by the PV plant.

\section{Post-processing methods and forecast evaluation}
\label{sec3}
In what follows, denote by \ $f_1,f_2,\ldots ,f_{51} \in [0,1]$ \ the 51-member normalized PV energy ensemble forecast of a given forecast horizon for a given PV power plant and time point, where \ $f_1=f_{\text{CTRL}}$ \ stands for the control forecasts, while \ $f_2,f_3, \ldots ,f_{51}$ \ are the 50  statistically indistinguishable, hence exchangeable ensemble members, referred to also as \ $f_{\text{ENS},1}, f_{\text{ENS},2}, \ldots , f_{\text{ENS},50}$. \ Furthermore denote by \ $\overline f_{\text{ENS}}$ \ the mean, and by \ $S^2_{\text{ENS}}$ \ the variance of these 50 exchangeable ensemble members, that is
\begin{equation*}
  \overline f_{\text{ENS}}:=\frac 1{50}\sum_{k=1}^{50}f_{\text{ENS},k} \qquad
  \text{and} \qquad S^2_{\text{ENS}}:=\frac 1{49}\sum_{k=1}^{50} \big(f_{\text{ENS},k} - \overline f_{\text{ENS}}\big)^2.
\end{equation*}

\subsection{Parametric methods}
\label{subs3.1}
As mentioned in the Introduction, parametric post-processing methods result in full predictive distributions, and in the EMOS and DRN approaches building on a single parametric law, the chosen distribution family strongly depends on the properties of the predictable quantity. Temperature is usually considered Gaussian \citep[see e.g.][]{grwg05,rl18}, wind speed requires a non-negative and skewed distribution such as truncated normal \citep{tg10} or log-normal \citep{bl15}, while the positive probability of observing zero precipitation can be handled by left-censoring a skewed distribution such as generalized extreme value \citep{sch14} or shifted gamma \citep{bn16} from below at zero.
The same idea led to parametric post-processing models for solar irradiance based on left-censored logistic or Gaussian \citep[see e.g.][]{sealb21,bb24} laws. However, in the case of PV power, the support of the predictive distribution has a natural upper bound induced by the maximal capacity of the given solar plant. 
Following  \citet{hkl25}, the predictive distribution in the EMOS and DRN models detailed in Sections \ref{subs3.1.1} and \ref{subs3.1.2}, respectively, follows a doubly censored Gaussian assigning point masses to zero and one, i.e., to both ends of the interval of possible (normalized) PV power values.

\subsubsection{Ensemble model output statistics}
\label{subs3.1.1}
Consider a Gaussian distribution \ $\mathcal N_0^1\big(\mu,\sigma^2\big)$ \ with location \ $\mu$ \ and scale \ $\sigma$ \ left-censored at zero and right-censored at one, characterized by the cumulative distribution function (CDF)
\begin{equation}
  \label{eq:cnormCDF}
  G\big(x\vert \mu,\sigma\big):=\begin{cases}
    0, & x <0, \\ 
    \Phi \left(\frac{x - \mu}\sigma\right), & 0\leq x \leq 1, \\
    1, & x>1,
    \end{cases}
\end{equation}
where \ $\Phi$ \ denotes the CDF of the standard Gaussian law. This distribution assigns masses \ $p_{\text{LB}}:=G\big(0\vert \mu,\sigma\big)= \Phi\big(-\mu/\sigma \big)$ \ to the origin and \ $p_{\text{UB}}:= 1-G\big(1\vert\mu,\sigma\big)=\Phi\big((\mu -1)/\sigma\big)$ \ to one, while the $p$-quantile \ $q_p, \ (0<p<1)$ \ of \eqref{eq:cnormCDF} equals \ $0$, \ if \ $p\leq p_{\text{LB}}$, \ the solution of \ $G\big(q_p\vert \mu,\sigma\big)=p$, \ if \ $p_{\text{LB}} <p < 1 - p_{\text{UB}}$, \ and \ $1$, \ if \ $p\geq 1-p_{\text{UB}}$. \

The parameters of our censored normal (CN) EMOS predictive distribution for (normed) PV power are linked to the (normed) ensemble members via equations
\begin{equation}
  \label{eq:CN_EMOS}
\mu = \alpha_0 + \alpha_1 f_{\text{CTRL}} + \alpha_2  \overline f_{\text{ENS}} \qquad \text{and} \qquad \sigma = \exp\left(\beta_0 + \beta_1 \log S_{\text{ENS}}^2\right).
\end{equation}
Following the optimum score approach suggested by \citet{gr07}, model parameters \ $\alpha_0,\alpha_1,\alpha_2,\beta_0,\beta_1 \in{\mathbb R}$ \ are estimated by optimizing the mean value of a proper verification score over training data comprising past forecast-observation pairs. The most popular choices are the ignorance score, which is the negative logarithm of the predictive probability density function (PDF) evaluated at the verifying observation \citep[see, e.g.,][Section 9.5.3]{w19}, which leads to the maximum likelihood estimates, and the continuous ranked probability score (CRPS), defined in Section \ref{subs3.3}. In the case study of Section \ref{sec4}, we utilize the latter.

While EMOS models give a computationally simple yet powerful tool for statistical post-processing, the rigid functional form of the link functions connecting the ensemble forecasts to the distributional parameters generally does not provide a straightforward way of including additional covariates such as forecasts of related weather quantities or location-specific data like geographical coordinates, altitude, or land use. Moreover, too many predictors might easily lead to overfitting, deteriorating the forecast performance. To circumvent this problem, \citet{mmz17} introduced a boosting algorithm that automatically selects the most important predictors in a nonhomogeneous regression model and provides the maximum likelihood estimates of the corresponding parameters. An implementation of the proposed approach for censored Gaussian predictive distribution can be found in the {\tt R} package {\tt crch} \citep{mmz16}.

\subsubsection{Distributional regression network}
\label{subs3.1.2}
Distributional regression networks (DRN), introduced in \citet{rl18}, provide an estimation of the parameters of the doubly censored normal predictive distribution by optimizing the mean of the corresponding CRPS over the training data as the loss function of a feedforward neural network. In contrast to EMOS models, considering an extended set of input variables with additional features is straightforward. Further, the large variability of the possible network structures and hyperparameters enables a more flexible post-processing method where non-linear relations are learned in an automated, data-driven manner. A drawback of the DRN models is that, due to the large number of weights, the training typically requires a larger amount of training data. However, in the present case study, the training period is long enough to use the same data as in the case of the EMOS models; for details, see Section \ref{subs4.1}.

\subsection{Non-parametric approaches}
\label{subs3.2}
The non-parametric post-processing methods considered here represent the predictive CDF \ $F$ \ by its \ $\tau$-quantiles, \ $q_\tau(F) := F^{-1}(\tau):= \inf\{y:F(y)\geq\tau\}$, \ which are estimated by quantile regression (QR). 
The most widely used loss function for quantile regression is an asymmetric linear loss called the pinball or quantile loss, which is defined as 
\begin{equation}
    \label{eq:pinball}
\rho_\tau(u) := 
\begin{cases}
	u \tau,  & \text{if  $u \geq 0$,} \\
		u (\tau-1), & \text{if  $u < 0$,}
        \end{cases}
\end{equation}
and minimized by \ $q_\tau(F)$. \ The quantile loss is not differentiable at \ $u = 0$, \ which may cause issues with convergence in the gradient-based optimization methods used for training neural networks. A remedy is to use the quantile Huber loss \citep{c11hub}, where the \ $u$ \ in \eqref{eq:pinball} is replaced by the Huber norm \ $h(u),$ \  calculated as
\begin{equation}
    \label{eq:huber}
h(u) := 
\begin{cases}
	\frac{u^2}{2\epsilon} ,  & \text{if  $|u| \leq \epsilon$,} \\
	|u| - \frac{\epsilon}{2}, & \text{if  $|u| > \epsilon$.}
        \end{cases}
\end{equation}
If a small value is selected for the \ $\epsilon$ \ threshold, e.g. \ $\epsilon = 10^{-8}$, \ the quantile Huber loss function closely approximates the quantile loss while being differentiable everywhere.

\subsubsection{Linear quantile regression}
\label{subs3.2.1}
Linear quantile regression \citep[LQR;][]{koenker2005quantile} approximates the quantile \ $q_\tau$ \ as a linear combination of the predictors. For the calibration of the ensemble forecasts at hand, the predictors are the raw ensemble members, and the calibrated quantile is calculated as
\begin{equation}
    \label{eq:qrest}
    q_{\tau} = \beta_0 + \sum_{k=1}^{51}\beta_kf_k
\end{equation}
where \ $\beta_0, \beta_1, \ldots, \beta_{51} \in {\mathbb R}$ \ are the regression coefficients, fitted to minimize the quantile loss.

\subsubsection{Quantile regression neural network}
\label{subs3.2.2}
Quantile regression neural networks (QRNN) use a neural network to provide a nonlinear mapping between the predictors and the output quantiles \citep{t00qrnn}. The QRNN implemented in this study is a feed-forward multilayer perceptron neural network with 51 output neurons, each assigned a quantile loss function with different \ $\tau$ \ values. In this way, one single QRNN model can estimate all required quantiles simultaneously, eliminating the need for training separate models for each quantile, in contrast to the LQR. 

To ensure the optimal fit of the model, an early stopping routine is applied. In this, a validation set is separated from the training data, which is not used directly to adjust the parameters of the model, but the loss function is evaluated for the validation set after each epoch. The training terminates when the validation loss stops improving for a pre-defined number of epochs called the early stopping patience. The convergence and accuracy of the QRNN highly depend on the selection of the hyperparameters, of which the most important are the number of hidden layers and the number of neurons in each hidden layer, the activation function, the learning rate, the early stopping patience, and the batch size.

\subsubsection{Bernstein quantile network}
\label{subs3.2.3}
Bernstein quantile networks (BQN), proposed by \citet{b20bqn}, estimate the whole quantile function as a Bernstein polynomial instead of individual quantiles. A Bernstein polynomial of degree \ $n$ \ is a linear combination of \ $n+1$ \ Bernstein basis polynomials, and the coefficients of the basis polynomials are calculated as the outputs of a neural network. The loss function of the training is the average quantile loss for a set of equidistant quantile levels. The degree of the Bernstein polynomial is a further hyperparameter of this method, in addition to those listed for the QRNN.

The method is further adjusted by constraining the coefficients to be nondecreasing, which implies that the quantile function is monotonically increasing \citep{sl22}. Technically, this is implemented by estimating the differences between the coefficients as non-negative values by using a softplus activation function in the output layer of the neural network. A monotonically increasing quantile function ensures that the forecasts for a higher quantile is always equal or higher than that of a lower quantile, i.e., \ $q_{\tau_1} \geq q_{\tau_2}$ \ for \ $\tau_1 > \tau_2$, \ in line with that by definition, a CDF must be monotonically increasing.

\subsubsection{Non-crossing quantile regression neural network}
\label{subs3.2.4}
Non-crossing quantile regression neural networks (NCQRNN), developed by \cite{sy24ncqrnn}, provide an alternative targeted solution to directly enforce the monotonicity of the CDF. QRNN is extended with an additional hidden layer before the output layer that ensures a non-decreasing mapping between the outputs of the previous layer and the nodes of the output layer. The main advantage of this approach is that it has no requirements on the network structure before the non-crossing layer, and thus it can be integrated into any type of neural network. However, NCQRNN has shown a decent performance even with a multilayer perceptron before the non-crossing layer in \citet{sy24ncqrnn}, therefore, this structure is used in the present study. An additional hyperparameter of NCQRNN over the QRNN is the number of neurons in the non-crossing layer, which must be equal to or higher than the number of output neurons.

\subsection{Forecast evaluation}
\label{subs3.3}
The performance of both probabilistic predictions (a forecast ensemble, a full predictive distribution, or predictive quantiles) and point forecasts (median or mean of the forecast ensemble or predictive distribution) can be evaluated with the help of scoring rules, which are loss functions assigning numerical values to forecast-observation pairs. In the case of the predictive median, we consider the mean absolute error (MAE), while the mean forecasts are evaluated with the help of the root mean squared error (RMSE) and the mean bias error (MBE), also known as the mean error \citep[see e.g.][Section 9.3.1]{w19}. Since the MAE is minimized by the median and the mean squared error is minimized by the mean \citep{g11det}, the aforementioned pairing of the point forecast with the error metrics ensures the consistency of the verification, i.e., the deterministic forecasts are only evaluated with metrics that they are optimal for \citep{k20, my23cal}.

In the case of probabilistic forecasts, one of the most popular scoring rules is the continuous ranked probability score \citep[CRPS;][Section 9.5.1]{w19}, as it is strictly proper and simultaneously addresses the calibration and the sharpness of the prediction \citep{gr07}. Calibration refers to statistical consistency between the probabilistic forecast and the corresponding observation, while sharpness refers to the concentration of the prediction.
When a probabilistic forecast corresponding to an observation \ $x\in {\mathbb R}$ \ materializes in the form of a predictive CDF \ $F$, \ the CRPS is defined as
\begin{equation}
    \label{eq:CRPSdef}
\crps(F,x) := \int_{-\infty}^{\infty}\Big[F(y)-{\mathbb I}_{\{y\geq x\}}\Big]^2{\mathrm d}y ={\mathsf E}|X-x|-\frac 12
{\mathsf E}|X-X'|,
\end{equation}
where \ ${\mathbb I}_A$ \ denotes the indicator function of a set \ $A$, \ while \ $X$ \ and \ $X'$ \ are independent random variables distributed according to \ $F$ \ and having a finite first moment. Note that for the doubly censored Gaussian distribution, the CRPS has a closed form \citep{jkl19} that allows efficient estimation of the parameters of the EMOS model presented in Section \ref{subs3.1.1}.

In the case of a forecast ensemble \ $f_1,f_2, \ldots ,f_K$, \ in \eqref{eq:CRPSdef} the predictive CDF \ $F$ \ should be replaced by the empirical CDF \ $\widehat F_K$, \ resulting in the expression
\begin{equation}
  \label{eq:ensCRPSdef}
\crps(\widehat F_K,x)=\frac 1K\sum_{k=1}^K\big |f_k -x \big| - \frac 1{2K^2} \sum_{k=1}^K\sum_{\ell =1}^K\big | f_k - f_{\ell}\big |, 
\end{equation}
see, e.g., \citet{kltg21}. This version of the empirical CRPS is implemented in the {\tt scoringRules} package of {\tt R} \citep{jkl19} and slightly differs from the ensemble CRPS defined in \citet[][Section 9.7.3]{w19}. The same formula \eqref{eq:ensCRPSdef} for the CRPS also applies when the predictive distribution is represented by its quantiles.

Furthermore, similar to other strictly proper scoring rules, the CRPS has an algebraic decomposition into a reliability (REL), resolution (RES) and uncertainty (UNC) term
\begin{equation*}
    \crps = \rel - \res + \unc,
\end{equation*}
where reliability summarizes the calibration of the probabilistic forecast, resolution is closely related to its sharpness, while uncertainty represents the climatological variability and thus depends only on observations \citep{h00,b09}.

In Section \ref{sec4}, the predictive performance of a forecast \ $F$ \ for a given time of the day is quantified, among others, with the help of the mean CRPS and the MAE over all forecast cases used for verification. For ranking the different predictions, we also consider the continuous ranked probability skill score \citep[CRPSS; see e.g.][]{gr07} and the mean absolute error skill score (MAES), which provide the improvement in mean CRPS and MAE of a forecast \ $F$ \ over a reference forecast \ $F_{\text{ref}}$, \ and are defined as
\begin{equation*}
    \label{eq:CRPSSdef}
   \crpss := 1 - \frac{\overline\crps_F}{\overline\crps_{F_{\text{ref}}}} \qquad \text{and} \qquad \maes:= 1- \frac{\mae_F}{\mae_{F_\text{ref}}},
 \end{equation*}
 where \ $\overline\crps_F, \ \mae_F$ \ and \ $\overline\crps_{F_{\text{ref}}}, \ \mae_{F_\text{ref}}$ \ denote mean score values corresponding to forecasts \ $F$ \ and \ $F_{\text{ref}}$, \ respectively.

Furthermore, to get an insight into the forecast skill of the quantile forecasts, we make use of the quantile score \citep[QS;][Section 9.6.1]{w19}, defined via the pinball loss \eqref{eq:pinball} as 
\begin{equation}
    \label{eq:QSdef}
\qs _{\tau}(F,x):=\rho_\tau\big(x -q_{\tau} (F) \big).
\end{equation}
Note that the QS is proper and its integral over all quantiles results in half of the CRPS \citep{gr11}.

Calibration and sharpness of predictive distributions can also be assessed with the help of the coverage and average width of \ $(1-\alpha)100\,\%, \ \alpha \in (0,1),$ \ central prediction intervals (intervals between the lower and upper \ $\alpha/2$ \ quantiles of the predictive CDF), respectively. In this context, prediction interval coverage probability (PICP) is the proportion of verifying observations located in the corresponding central prediction interval, which for a calibrated forecast should be around \ $(1-\alpha)100\,\%$, \ while the prediction interval average width (PIAW) quantifies the concentration of the predictive law. Note that when a $K$-member ensemble forecast is also involved in the study, to ensure fair comparability in the detail analysis, level \ $\alpha$ \ is chosen to match its nominal coverage of \ $(K-1)/(K+1)100\,\%$ \ meaning \ $96.15\,\%$ \ for the 51-member PV forecasts at hand. Moreover, since the prediction interval of interest depends on the application of the forecasts, PIAW is also presented for all possible central prediction intervals defined by the 51-member ensemble as a function of the nominal and empirical coverage rates.

However, PICP alone is not sufficient to evaluate the reliability of probabilistic forecasts, since it may suggest perfect reliability even if both quantiles defining the prediction interval are biased in the same direction \citep{ldp19}. A better approach is to evaluate the reliability at all quantile levels individually using a reliability diagram that plots the proportion of the observations that are actually smaller than the forecasts for each quantile at hand as a function of the quantile level. The reliability curve of a perfectly calibrated forecast lies close to the diagonal.

Another simple graphical tool for visual assessment of calibration of probabilistic forecasts given either as a forecast ensemble or as a sample drawn from a predictive distribution is the verification rank histogram \citep[][Section 9.7.1]{w19}. The verification rank is defined as the rank of the observation with respect to the corresponding forecast, which for a calibrated $K$-member ensemble should be uniformly distributed on the set \ $\{1,2, \ldots, K+1\}$. \ Bias results in triangular shapes, while $\cup$- and $\cap$-shaped rank histograms suggest under- and overdispersion. Moreover, one can also quantify the deviation from the uniform distribution with the help of the reliability index \citep[RI;][]{dmhzds06}, defined as
 \begin{equation*}
   \label{eq:relind}
 \ri:=\sum_{r=1}^{K+1}\Big| \rho_r-\frac 1{K+1}\Big|,
\end{equation*}
where \ $\rho_r$ \ is the relative frequency of rank \ $r$ \ over all forecast cases in the verification period.

\section{Results}
\label{sec4}
In the following, we present a detailed comparison of the predictive performance of parametric and non-parametric post-processing methods introduced in Sections \ref{subs3.1} and \ref{subs3.2}, respectively. All models are trained locally; i.e., post-processing models for a given PV power plant are merely based on past forecast-observation pairs for that very location. For validation, we use power data for calendar year 2023 and, as mentioned in Section \ref{subs2.1}, consider only daytime predictions corresponding to positive observed PV power, meaning at most 65 observations/day between 03:00 and 19:00 UTC. Both the parametric and non-parametric post-processing methods are based on normalized data, i.e., both PV power forecasts and PV power outputs of a plant are normed by the corresponding nominal AC power provided in Table \ref{tab:locations}. 
Furthermore, to ensure a fair comparability of the parametric methods resulting in full predictive distributions, non-parametric techniques providing quantile forecasts, and the raw ensemble, we consider 51 equidistant quantiles from the predictive distributions for each post-processing model. These quantiles are then transformed back to the original scale. Finally, the differences between the nominal powers of the seven considered plans are compensated for by reporting score values normalized by the mean daytime power output of the PV plans listed in Table \ref{tab:meanPower}.

\begin{table}[t]  
  \begin{center}{\footnotesize
    \begin{tabular}{l|ccccccc}
       \hline
      Name&Bodajk&Cegl\'ed&Fels\H ozsolca&Fert\H osz\'eplak&Magyarsarl\'os&Paks&\'Ujk\'{\i}gy\'os
\\ \hline
 Power (kW)&169.27& 178.29&5503.95&168.37&166.27&6161.85&179.06\\ \hline      
       \end{tabular}
       }
    \end{center}
    \caption{Mean daytime power output of the PV plants.}
    \label{tab:meanPower}
\end{table}

\subsection{Implementation details}
\label{subs4.1}
All 51 members of the ECMWF ensemble forecasts contain GHI, ambient temperature, and wind speed data, which are converted to PV power using a physical model chain. All perturbed NWP ensemble members are generated from randomly issued initial conditions; therefore, there is no continuity between the same-numbered members of different model runs. To that end, the 50 perturbed PV power forecast members are sorted in ascending order in each timestep, which can be seen as converting them to equidistant quantile forecasts, and the sorted ensemble is used as the input of the post-processing models.

In the case of EMOS modelling, all lead times are considered separately, leading to at most 65 distinct models per PV power plant. We consider a fixed training period of 1460 calendar days between 1 January 2019 and 30 December 2022. Note that a rolling training window of the same length has also been tested without providing significantly different results, whereas shorter training periods (365-, 730-, 1096-day have been tested) decrease the forecast skill. As mentioned in Section \ref{subs3.1.1}, the parameters of the doubly censored normal EMOS (CN EMOS) model are estimated by optimizing the mean CRPS over the training data, while in the boosted version of EMOS, referred to as CN EMOS-B, the control member \ $f_{\text{CTRL}}$ \ and the mean \ $\overline f_{\text{ENS}}$ \ and standard deviation  \ $S_{\text{ENS}}$ \ of the exchangeable ensemble members are used as covariates.

For CN EMOS-B and all the other post-processing methods, we use the same training period as for the EMOS model, but all the lead times are pooled, resulting in a single trained model for each method and PV plant.

\begin{table}[t]  
  \begin{center}{\footnotesize
    \begin{tabular}{l|cccc|c|cc}
       \hline
       &\multicolumn{4}{|c|}{Probabilistic forecast}&Median&\multicolumn{2}{|c}{Mean} \\ \cline{2-8}
         Forecast &CRPS&Reliability&Resolution&CRPSS&MAE&MBE&RMSE \\ \hline
       CN EMOS&18.95\%&1.66\%&32.54\%&11.13\%&26.46\%&5.19\%&42.41\% \\
CN EMOS–B&18.96\%&1.53\%&32.40\%&11.08\%&26.22\%&4.16\%&42.20\% \\
CN DRN&18.58\%&1.97\%&{\bf 33.22\%}&12.85\%&26.00\%&3.48\%&42.14\% \\ \hline
LQR&18.65\%&0.46\%&31.64\%&12.56\%&26.39\%&2.12\%&42.03\% \\
QRNN&{\bf 18.18\%}&{\bf 0.44\%}&32.09\%&{\bf 14.73\%}&25.84\%&1.68\%&{\bf 41.86\%} \\
BQN&18.19\%&0.83\%&32.48\%&14.69\%&{\bf 25.78\%}&2.07\%&41.91\% \\
NCQRNN&18.20\%&0.47\%&32.10\%&14.67\%&{\bf 25.78\%}&{\bf 1.65\%}&41.91\% \\ \hline
Ensemble&21.33\%&7.63\%&36.14\%&0.00\%&27.58\%&9.79\%&44.07\% \\ \hline
       \end{tabular}
       }
    \end{center}
    \caption{Summary CRPS, reliability, and resolution of the probabilistic forecasts and the MAE, MBE, and RMSE of the consistently summarized deterministic forecasts in average for all PV plants}
    \label{tab:scores}
\end{table}

For the doubly censored censored DRN (CN DRN) model, the neural network is a multilayer perceptron with three hidden layers consisting of 15, 10, and 10 neurons, respectively, all of which use a ReLU activation function. In the output layer, there are two neurons, corresponding to the number of estimated parameters. To ensure the non-negativity of the scale parameter, one of the neurons applies a softplus activation, while the other activation function is the linear one. The input features are simply the 51-member ensemble for the PV power forecast. To optimize the loss, we apply the Adam optimizer with a learning rate value of 0.001 and use a batch size of 256. An early stopping criterion terminates the training if the loss function value computed on a validation set does not decrease over six consecutive epochs. 
Following common practice in DRN-based post-processing, we train an ensemble of ten neural networks and average the output parameters \citep{schulz2022aggregating}. 

The QRNN, BQN, and NCQRNN models include a similar multilayer perceptron with up to two hidden layers with 5 to 200 neurons each. The considered activation functions are the ReLU, softplus, logistic, and tanh functions, the learning rate is selected from the 0.0005 to 0.05 interval, the early stopping patience may range from 5 to 50 epochs, and the batch size is between 200 and 20000. The degree of the Bernstein polynomial in BQG ranges from 6 to 15, whereas the number of non-crossing neurons in NCQRNN is between 51 and 60. The optimal hyperparameters for each model and PV plant are selected from the aforementioned intervals/options using the Optuna framework \citep{a19optuna}. For this, the training data of four years is divided into five-day-long blocks, and the first three days of each block are used for the actual training of the models, the fourth days are used as validation data for the early stopping, and the fifth days are used to form a holdout dataset. The Optuna hyperparameter optimization studies were run with 100 trials for each model, and the hyperparameter sets resulting in the lowest CRPS for the holdout data are selected. After the optimal hyperparameters are found, the data for every fifth day is used for validation, and the rest for the training of the final model.

\begin{table}[t]  
  \begin{center}{\footnotesize
    \begin{tabular}{l|ccccccc}
       \hline
      Forecast&Bodajk&Cegl\'ed&Fels\H ozsolca&Fert\H osz\'eplak&Magyarsarl\'os&Paks&\'Ujk\'{\i}gy\'os
\\ \hline
       CN EMOS&19.44\%&19.01\%&20.62\%&19.83\%&17.57\%&17.77\%&18.42\% \\
CN EMOS–B&19.40\%&18.91\%&20.67\%&19.97\%&17.49\%&17.75\%&18.56\% \\
CN DRN&18.83\%&18.57\%&20.20\%&19.46\%&17.31\%&17.46\%&18.27\% \\ \hline
LQR&19.06\%&18.53\%&20.31\%&19.55\%&17.41\%&17.45\%&18.23\% \\
QRNN&{\bf 18.53\%}&18.13\%&19.67\%&{\bf 19.02\%}&17.10\%&17.07\%&{\bf 17.77\%} \\
BQN&{\bf 18.53\%}&{\bf 18.11\%}&{\bf 19.63\%}&19.05\%&17.14\%&{\bf 17.05\%}&17.84\% \\
NCQRNN&18.55\%&18.11\%&19.78\%&19.08\%&{\bf 17.01\%}&17.07\%&17.78\% \\ \hline
Ensemble&21.99\%&21.23\%&23.14\%&21.87\%&21.20\%&19.69\%&20.17\% \\ \hline
       \end{tabular}
       }
    \end{center}
    \caption{Overall mean CRPS of post-processed and raw PV power forecasts normalized to the
mean daytime power output of the PV plants.}
    \label{tab:crps}
\end{table}

\subsection{Performance of probabilistic PV forecasts}
\label{subs4.2}
As mentioned, all reported scores are normalized by the mean daytime power output of the corresponding PV plant, hence expressed in percentage of the normalizing constant.

\begin{figure}[t]
\begin{center}
\epsfig{file=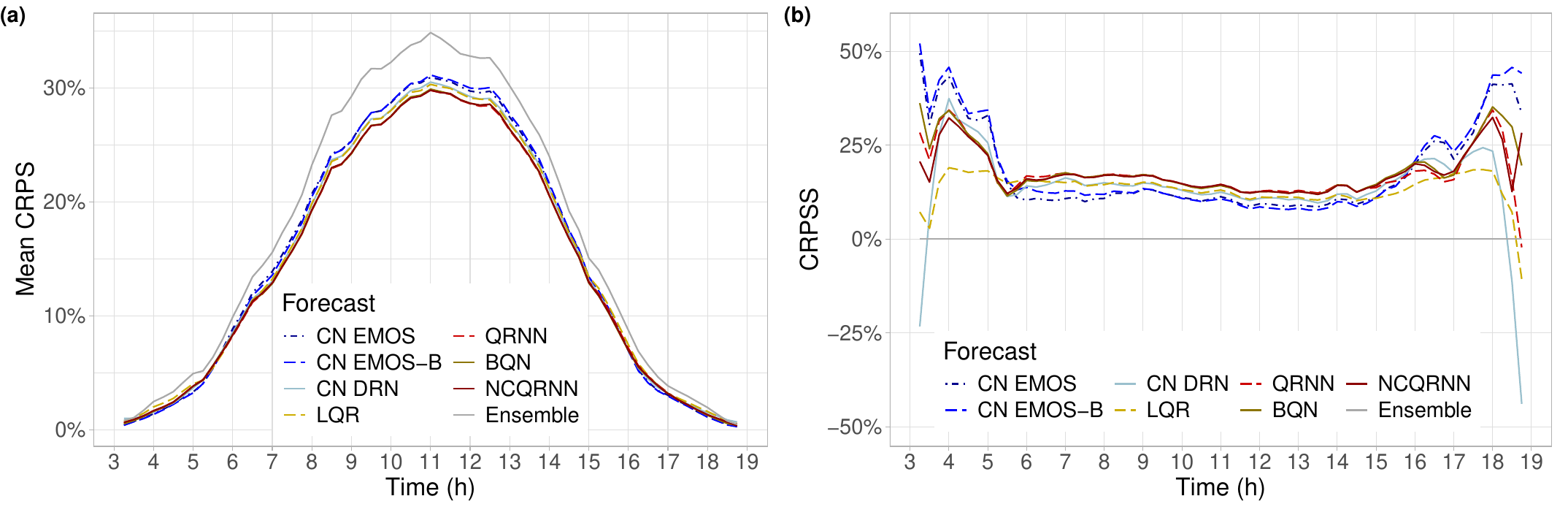, width=\textwidth}
\end{center}
\caption{Mean CRPS of post-processed and raw PV power forecasts normalized to the mean daytime power output of the PV plants (a) and CRPSS of post-processed forecasts with respect to the raw ensemble (b) as functions of the observation time.}
\label{fig:crps_crpss}
\end{figure}

For an overall evaluation, consider first the mean scores averaged over all PV plants in Table \ref{tab:scores} for all post-processed and raw PV forecasts. The lowest CRPS is achieved by the nonlinear QR methods, namely the QRNN, BQN, and NCQRNN, achieving a CRPS reduction of 14.67 -- 14.73\,\% over the raw ensemble (represented by the CRPSS); however, even the least effective EMOS models reach a CRPSS of 11.08\,\%. The reliability-resolution (REL-RES) decomposition reveals that all methods improve the reliability of the forecast at the cost of a decreased resolution (sharpness). Even though all methods are able to improve the reliability substantially, the non-parametric models achieve better reliability values of 0.44 -- 0.83\,\% compared to 1.53 -- 1.97\,\% of the parametric models. However, the less effective calibration of the parametric models is partly compensated by a slightly higher resolution.

To address the dependence of the results on the PV plant locations, the CRPS values are presented individually for each PV plant in Table \ref{tab:crps}. The conclusions drawn from the mean score values also hold for all PV plants, as there are no significant differences between the order of the methods. The lowest CRPS is everywhere achieved by one of the nonlinear QR models, which ended up head-to-head in all locations, with each being the best performer in at least one PV plant. The achieved CRPSS of the best model, however, strongly depends on the location, with the lowest and highest CRPSS values being 11.89\,\% and 19.32\,\% in \'Ujk{\'\i}gy\'os and Magyarsarl\'os, respectively.

\begin{figure}[t]
\begin{center}
\epsfig{file=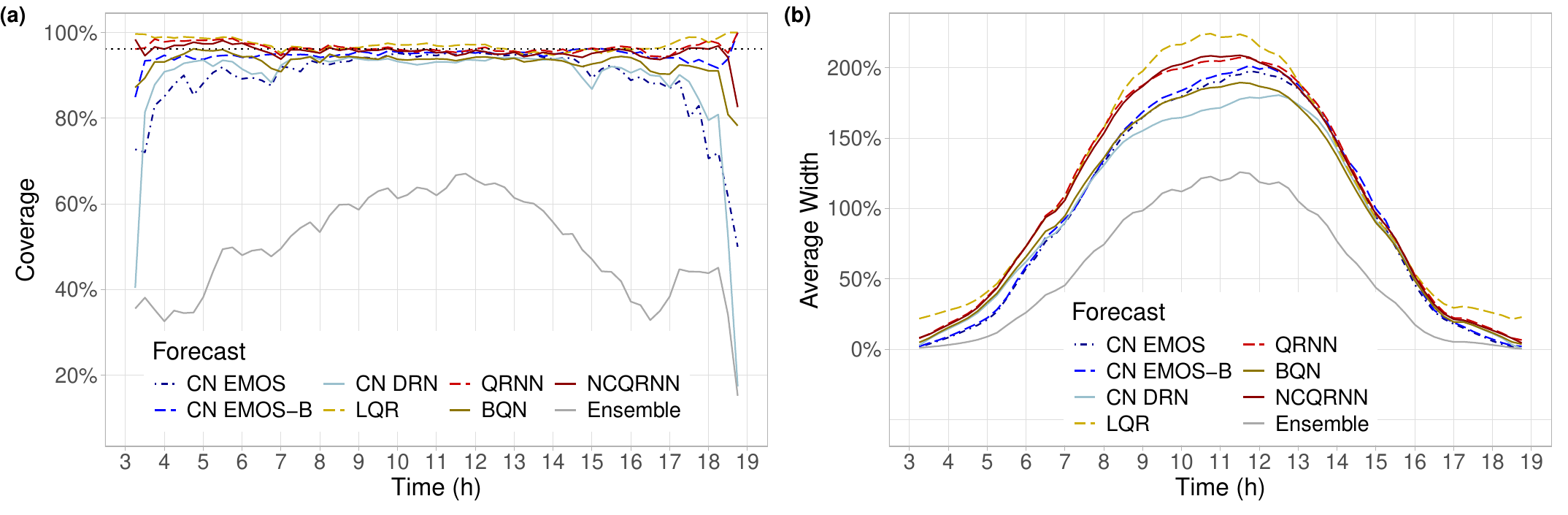, width=\textwidth}
\end{center}
\caption{Coverage (PICP) (a) and average width (PIAW) (b) of nominal $96.15\,\%$ central prediction intervals of post-processed and raw PV power forecasts normalized to the mean daytime power output of the PV plants as functions of the observation time. In panel (a), the ideal coverage is indicated by the horizontal dotted line.}
\label{fig:cov_aw}
\end{figure}

The mean CRPS of post-processed and raw PV forecasts for each hour of the day are displayed in Figure \ref{fig:crps_crpss}a. As confirmed by the skill scores in Figure \ref{fig:crps_crpss}b, 
compared to the raw ensemble, post-processing results in a substantial relative improvement of around 10\,\% during the hours of peak PV power production (06:00 -- 16:00 UTC). In this period of the day, the differences between the competing calibrated forecasts are rather small with the advanced quantile-based methods (QRNN, BQN, NCQRNN) exhibiting the best, almost identical skill, followed by the LQR and CN DRN, whereas the two EMOS variants are slightly behind. This trend is in line with the conclusions drawn from the overall scores of Tables \ref{tab:scores} and \ref{tab:crps}. Results on the significance of the differences in mean CRPS among the various forecasts are presented in the Appendix. They confirm that all post-processing methods significantly outperform the raw PV power ensemble, and the advantage of the three quantile-based methods that perform the best over the other four approaches is also significant at a 5\,\% level.

\begin{figure}[t]
\begin{center}
\epsfig{file=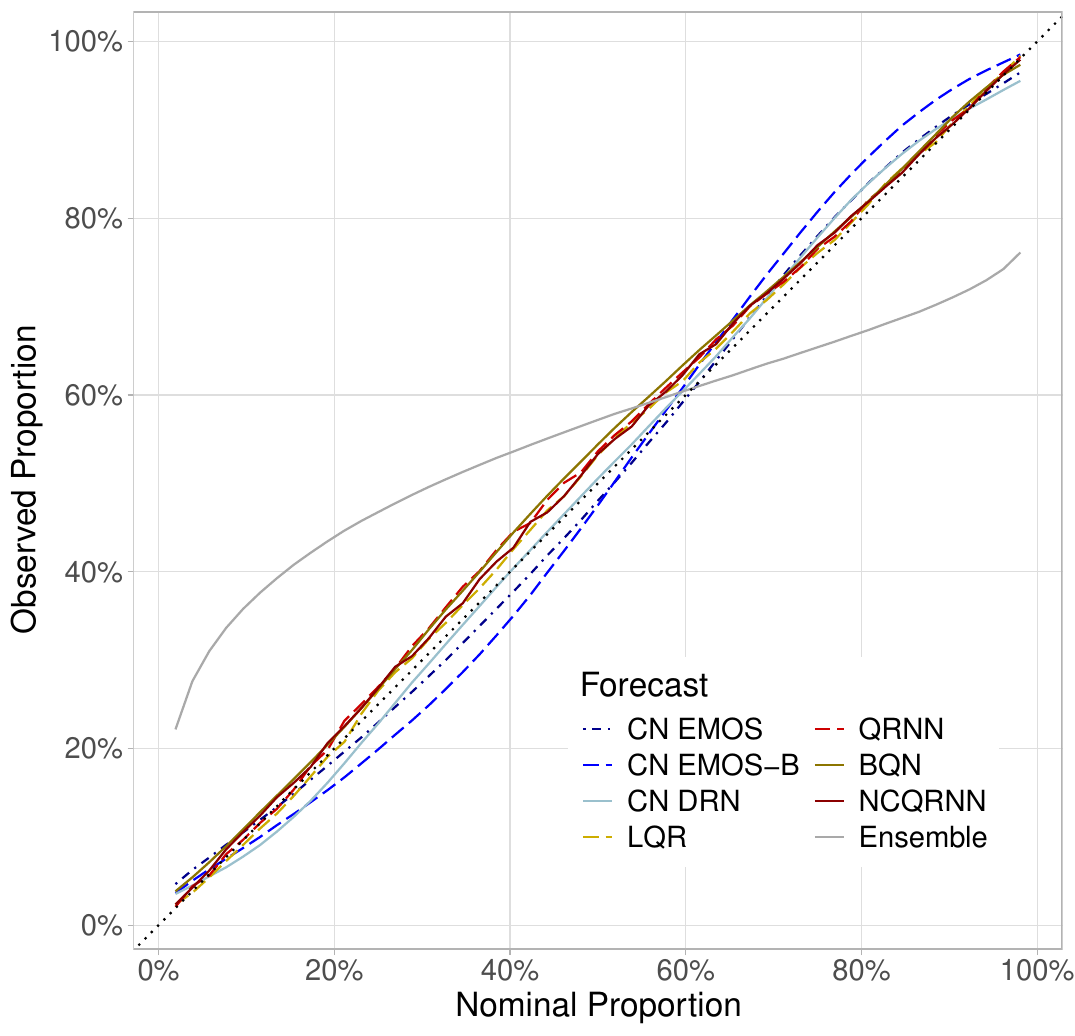, width=.5\textwidth}
\end{center}
\caption{Reliability diagrams of post-processed and raw PV power forecasts.}
\label{fig:reldiag}
\end{figure}

\begin{figure}[h]
\begin{center}
\epsfig{file=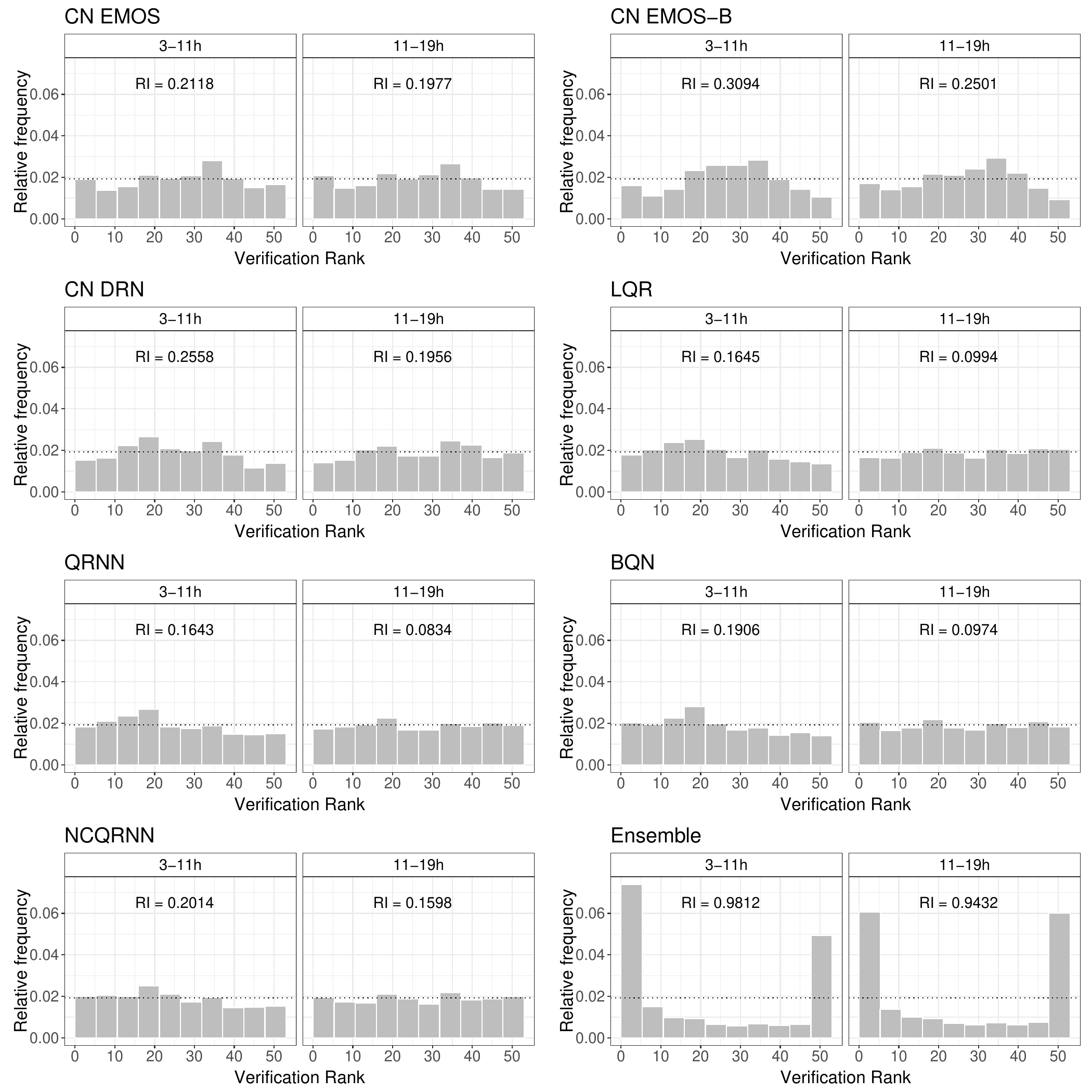, width=.85\textwidth}
\end{center}
\caption{Verification rank histograms of post-processed and raw PV power forecasts together with the corresponding reliability indices for observation times 3–11 h and 11–19 h.}
\label{fig:pit}
\end{figure}

The improved calibration of post-processed forecasts is also clearly visible in the coverage (PICP) values presented in Figure \ref{fig:cov_aw}a. While the maximal PICP of the raw forecasts is just slightly above 67\,\%, between 08:00 -- 14:00 UTC, all post-processing approaches result in almost perfect coverage, which is maintained by the best-performing non-parametric approaches for all observation times, but the most extreme ones. The reliability diagram in Figure \ref{fig:reldiag} and the rank histograms in Figure \ref{fig:pit} allow a more detailed assessment of reliability over the whole range of probability levels. These diagrams not only confirm the significant underdispersion of the raw ensemble but also reveal that both versions of the EMOS model overcompensate this, turning into a slight overdispersion in the medium probability range.

\begin{figure}[t]
\begin{center}
\epsfig{file=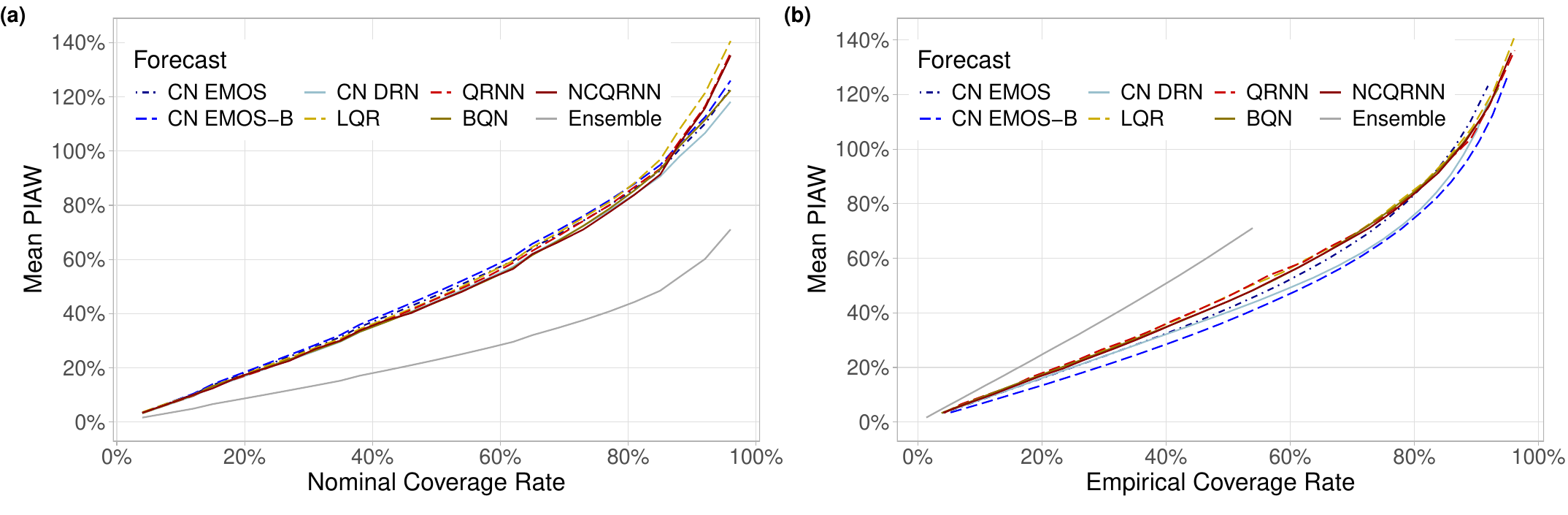, width=\textwidth}
\end{center}
\caption{Sharpness of post-processed and raw PV power forecasts normalized to the mean daytime power output of the PV plants as functions of the nominal coverage (a) and the corresponding empirical coverage (PICP) (b).}
\label{fig:shp}
\end{figure}

As indicated in Figure \ref{fig:cov_aw}b, the price of the better calibration is the loss in sharpness. Among the competing calibrated forecasts, the CN DRN results in the lowest overall PIAW, followed by the BQN and the two EMOS methods. Figure \ref{fig:shp}a shows the mean PIAW for all central prediction intervals with different nominal coverage rates, clearly revealing the widening of all prediction intervals as a result of the calibration. That said, this diagram does not account for the fact that the prediction intervals of an uncalibrated ensemble cover a significantly lower proportion of the observation as compared to what their nominal coverage rate suggests, and thus misleadingly imply the deterioration of the forecast quality. To better represent the sharpness of the forecasts with respect to their calibration, Figure \ref{fig:shp}b plots the mean PIAW as a function of the empirical coverage rate (i.e., the PICP) instead of the nominal one. This novel graphical representation reveals that, compared to the proportion of the observations they cover, the calibrated prediction intervals are narrower compared to the raw ensemble, in line with the improved CRPS.

\begin{figure}[t]
\begin{center}
\epsfig{file=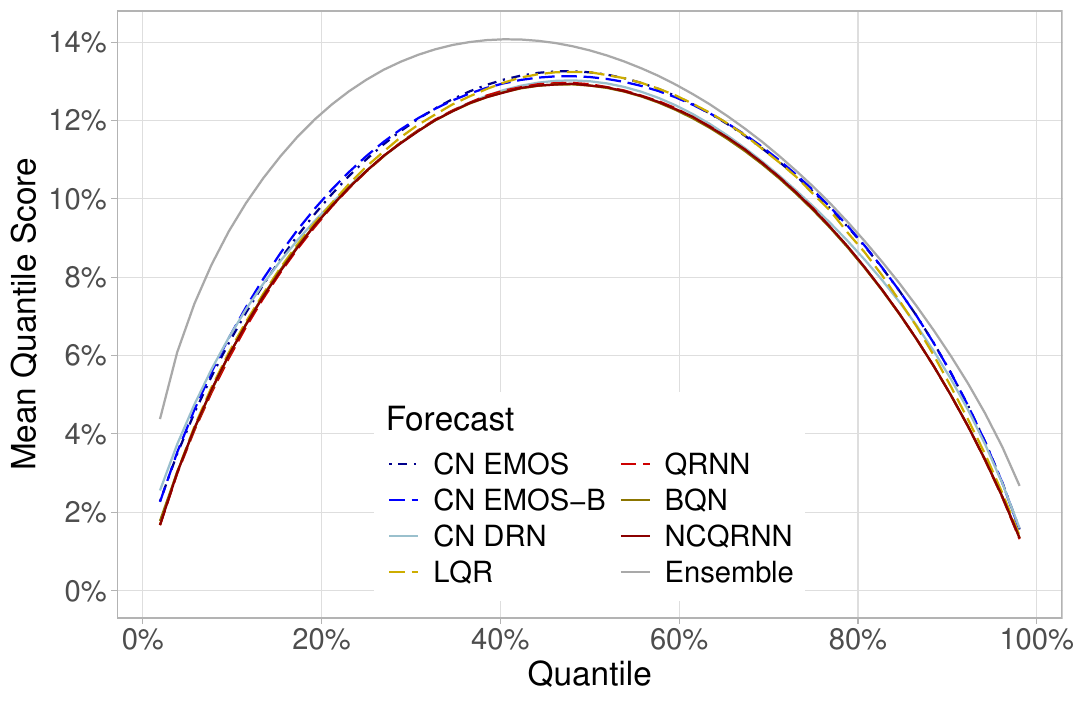, width=.5\textwidth}
\end{center}
\caption{Quantile score of post-processed and raw PV power forecasts normalized to the mean daytime power output of the PV plants.}
\label{fig:pin}
\end{figure}

Considering that the CRPS is double the integral of the QS over all quantiles \citep{gr11, b12}, the QS diagram in Figure \ref{fig:pin} can reveal which quantile levels contribute the most to the CRPS improvement. Since, according to the reliability diagram in Figure \ref{fig:reldiag}, the raw ensemble is only reliably around the 62\,\% probability level, the biggest improvement in QS is achieved at the lower quantile range. Comparing the different post-processing methods, the QRNN, BQN, and NCQRNN have almost the same QS for all quantiles, which is slightly but consistently lower than the QS of other methods. The LQR performs well up to the 30\,\% and above the 90\,\% quantile, but lags behind in between. In contrast, the DRN catches up with the nonlinear QR methods between 20\,\% and 70\,\%, but slightly underperforms around the extreme probability levels.

In terms of the accuracy of point forecasts, the average error metrics for all PV plants are presented in Table \ref{tab:scores}. The greatest relative MAE improvement over the raw ensemble is 6.51\,\%, achieved by either the BQN or NCQRNN models, whereas the greatest RMSE improvement is 5.02\,\%, achieved by the QRNN. The raw forecasts have a significant positive MBE, originating from both the ensemble NWP and the model chain. All methods decrease the MBE, but the non-parametric approaches can provide a better correction as compared to the parametric models, among which the simple EMOS keeps more than half of the original bias.

\begin{figure}[t]
\begin{center}
\epsfig{file=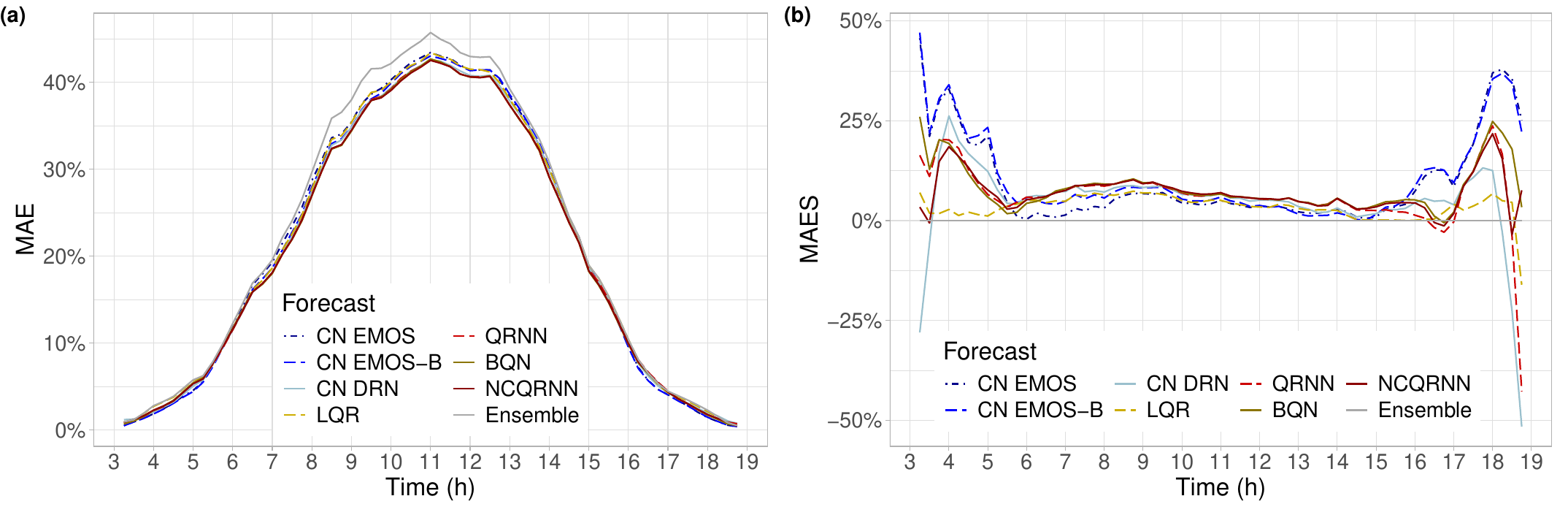, width=\textwidth}
\end{center}
\caption{MAE of the median of post-processed and raw PV power forecasts normalized to the mean daytime power output of the PV plants (a) and MAES of post-processed forecasts with respect to the raw ensemble (b) as functions of the observation time.}
\label{fig:mae_maes}
\end{figure}

According to Figure \ref{fig:mae_maes}, showing the MAE of the median of probabilistic forecasts and the MAES with respect to
the raw ensemble, while the ranking of the various predictions is slightly different, as CN DRN can often catch up with the best-performing non-parametric methods, the (normalized) improvement in peak hours (06:00 -- 16:00 UTC) hardly exceeds 12\,\%. For results addressing the significance of differences in MAE among the various forecasts, we again refer to the Appendix. A similar behaviour can be observed for the daily evolution of the RMSE of the mean forecasts (not shown). Overall, the gain of statistical post-processing in deterministic forecasting is not so striking, but still remarkable considering that the goal of the post-processing is probabilistic calibration and only the raw ensemble members are used as predictors. 

\begin{figure}[t]
\begin{center}
\epsfig{file=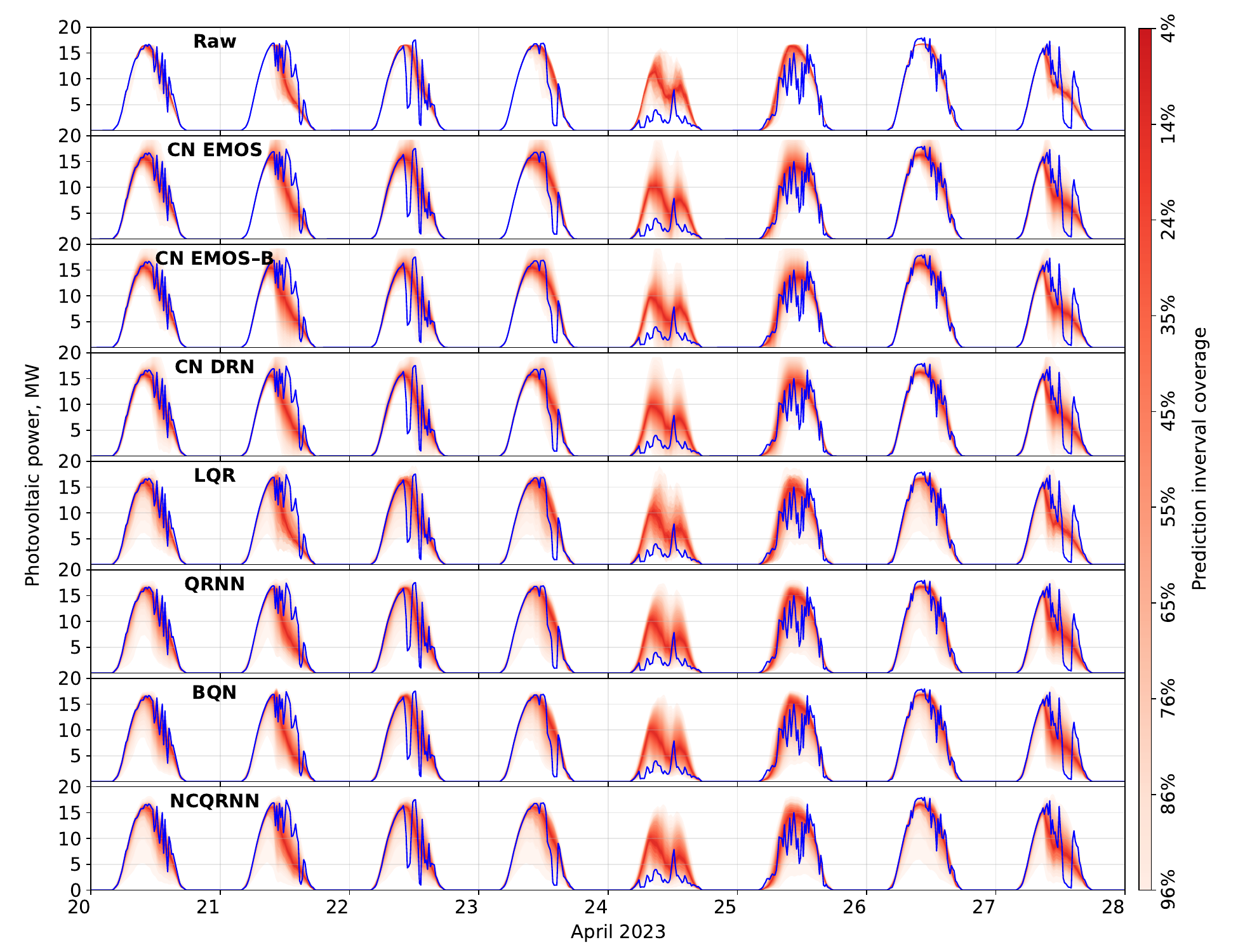, width=.95\textwidth}
\end{center}
\caption{Sample time series plot of the measured PV power production (blue) and the of the raw and calibrated ensemble probabilistic forecasts (red) for the Paks PV plant}
\label{fig:timeseries}
\end{figure}

Finally, time series plots of the raw and all post-processed forecasts are shown for a sample week in Figure \ref{fig:timeseries}. There is no significant visual difference between the models within both the parametric and non-parametric model categories, but forecasts created by these two different approaches can be clearly distinguished. The main difference is that the parametric models assign narrow prediction intervals to the periods with supposedly clear sky (see the mornings of the 20th, 24th, and 27th of April), whereas the non-parametric methods assign much wider prediction intervals towards the lower PV power values, especially for the high coverage rates. In this way, these models give some probability to the events when the clear sky forecasts are wrong, which explains both the improved reliability and lower sharpness.

Overall, the results show that the nonlinear QR methods, namely the QRNN, BQN, and NCQRNN, are consistently the best performers in almost all respects. The lower performance of the LQR can be justified by its linearity, while for the parametric method, the pre-defined shape of the CDF limits the performance. Among the three nonlinear QR methods, the simple QRNN has a slight edge over the others, suggesting that the least constrained method can yield the best results in this application. However, one should note that a large historical dataset covering four full calendar years was available to train the models. When less data are available for training, constraints, for instance, on the CDF, can prove effective to avoid nonphysical results, and this can be the application where parametric methods excel.

\section{Conclusions}
\label{sec5}

The present work provides a detailed comparison of seven state-of-the-art approaches to statistical post-processing of 51-member PV power ensemble forecasts obtained from operational ECMWF ensemble weather forecasts as outputs of site-specific physical model chains. With the help of PV power data from seven PV plants in Hungary, we evaluated the skill of the doubly censored Gaussian ensemble model output statistics (CN EMOS) model, its boosted version (CN EMOS-B) and the corresponding distributional regression network (CN DRN) technique together with the linear quantile regression (LQR), quantile regression neural network (QRNN) and its non-crossing variant (NCQRNN) and Bernstein quantile network (BQN) methods. The chosen pool of post-processing methods represents, on the one hand, both parametric (CN EMOS, CN EMOS-B, CN DRN) and non-parametric techniques (LQR, QRNN, BQN, NCQRNN), and on the other hand, both traditional statistical (CN EMOS, CN EMOS-B, LQR) and machine learning-based approaches (CN DRN, QRNN, BQN, NCQRNN).

We found that compared to the raw PV power ensemble, any form of statistical post-processing significantly improves the predictive performance, resulting in, for instance, an 11.08 -- 14.73\,\% overall gain in terms of the mean continuous ranked probability score. Post-processing also decreases the quantile score, the mean absolute error of the median and the root mean squared error of the mean, improves the reliability, and yields almost perfect coverage of the nominal central prediction intervals; however, at the cost of a deterioration in sharpness.

From the competing post-processing methods, the advanced non-parametric quantile regression models (QRNN, BQN, and NCQRNN) behave very similarly, consistently outperforming the other four approaches, with the QRNN exhibiting the best overall skill. In general, non-parametric methods show superior predictive performance compared to the parametric models, which matches the results of, for instance, \citet{bmcdsnl25}; nevertheless, the best-performing parametric CN DRN can catch up with the least skillful non-parametric LQR. Furthermore, our study also confirms that machine learning-based approaches surpass their traditional statistical counterparts (CN DRN vs.\ CN EMOS or QRNN vs.\ LQR), even though here the full potential of neural networks stemming from their capability of easily accommodating additional relevant covariates was not exploited; all models relied merely on the same set of inputs given by PV power ensemble forecasts or their summary statistics.  

The post-processing methods considered in our study provide several avenues for further improvements and analysis.
For example, comparisons with alternative non-parametric approaches that are not directly based on quantile regression such as isotonic distributional regression \citep{henzi2021isotonic} or member-by-member post-processing \citep{van2015ensemble} might be of interest and could help to identify alternative approaches, although previous studies generally indicate a similar predictive performance to EMOS \citep{sl22}.
Further, it has been noted in the post-processing literature that a key reason of the success of modern machine learning-based methods is their capability to include additional covariates \citep[e.g.,][]{demaeyer_etal_2023_euppbench}.  
Therefore, the neural network-based parametric and non-parametric approaches might be further improved by extending their inputs with weather predictions from the NWP system. That said, \citet{hkl25} noted only minor improvements when doing so.
Another route towards improving the predictive performance might be a more sophisticated use of the lead time information, as for example proposed by \citet{Mlakar2024}. 
Further, instead of applying a single model chain only, it would also be possible to consider an ensemble of possible model chains \citep{my22mce}, leading to a multi-model ensemble forecast of PV power production.
Given the relevance of probabilistic energy forecasts is grid operations and electricity markets, the statistical evaluation based on scoring rules considered here should further be accompanied by considerations of other aspects, including economic impacts of improved forecasts \citep{gls23}.

Over the past years, there have been rapid advances in machine learning-based, purely data-driven weather models, including Pangu-Weather \citep{BiEtAl2023}, GraphCast \citep{LamEtAl2023}, or AIFS \citep{AIFS}, providing deterministic forecasts, and the more recent ensemble prediction systems like GenCast \citep{gencast25} or AIFS-CRPS \citep{lang2024aifs-crps}, which now outperform physics-based NWP models for a variety of weather variables. 

A key question in the context of solar energy forecasting is whether predictions from these data-driven weather models might replace NWP ensemble forecasts, and which role post-processing methods could play. For example, recent research has demonstrated that data-driven and physics-based weather models might equally benefit from post-processing \citep{bremnes2024evaluation,buelte_etal_2024}.
However, most of the current data-driven weather models do not provide relevant outputs such as GHI, although there have been relevant recent developments including the FuXi-2.0 model \citep{FuXi2}, which explicitly targets solar and wind energy forecasting.

\section*{Acknowledgments}
 The authors thank Norbert Péter from MVM Green Generation Ltd. for the PV plant design and production data and the HungaroMet Hungarian Meteorological Service for providing access to the ECMWF’s Meteorological Archival and Retrieval System. Martin J\'anos Mayer was supported by the National Research, Development and Innovation Fund under the project no. OTKA-FK 142702, and the Hungarian Academy of Sciences through the János Bolyai Research Scholarship. \'Agnes Baran and S\'andor Baran were supported by the Hungarian National Research, Development and Innovation Office under Grant K142849. The work leading to this paper was done, in part, during the visit of S\'andor Baran to the Heidelberg Institute for Theoretical Studies in July 2025 as a guest researcher. Sebastian Lerch and Nina Horat acknowledge funding from the Vector Stiftung within the Young Investigator Group ``Artificial Intelligence for Probabilistic Weather Forecasting''.

\bibliographystyle{rss}

\bibliography{PV_paper}

\begin{thebibliography}{82}
\expandafter\ifx\csname natexlab\endcsname\relax\def\natexlab#1{#1}\fi
\expandafter\ifx\csname url\endcsname\relax
  \def\url#1{\texttt{#1}}\fi
\expandafter\ifx\csname urlprefix\endcsname\relax\def\urlprefix{URL: }\fi

\bibitem[{Akiba et~al.(2019)Akiba, Sano, Yanase, Ohta and Koyama}]{a19optuna}
Akiba, T., Sano, S., Yanase, T., Ohta, T. and Koyama, M. (2019) Optuna: {{A
  Next-generation Hyperparameter Optimization Framework}}.
\newblock In \textit{Proceedings of the 25th {{ACM SIGKDD International
  Conference}} on {{Knowledge Discovery}} \& {{Data Mining}}}, {{KDD}} '19,
  2623--2631. New York, NY, USA: Association for Computing Machinery.

\bibitem[{Bakker et~al.(2019)Bakker, Whan, Knap and Schmeits}]{Bakker2019}
Bakker, K., Whan, K., Knap, W. and Schmeits, M. (2019) Comparison of
  statistical post-processing methods for probabilistic {NWP} forecasts of
  solar radiation.
\newblock \textit{Solar Energy}, \textbf{191}, 138–150.

\bibitem[{Baran and Baran(2024)}]{bb24}
Baran, {\'A}. and Baran, S. (2024) A two-step machine learning approach to
  statistical post-processing of weather forecasts for power generation.
\newblock \textit{Quarterly Journal of the Royal Meteorological Society},
  \textbf{150}, 1029--1047.

\bibitem[{Baran and Lerch(2015)}]{bl15}
Baran, S. and Lerch, S. (2015) Log-normal distribution based {EMOS} models for
  probabilistic wind speed forecasting.
\newblock \textit{Quarterly Journal of the Royal Meteorological Society},
  \textbf{141}, 2289--2299.

\bibitem[{Baran et~al.(2025)Baran, Mar\'{\i}n, Cuevas, D\'{\i}az, Szab\'o,
  Nicolis and Lakatos}]{bmcdsnl25}
Baran, S., Mar\'{\i}n, J.~C., Cuevas, O., D\'{\i}az, M., Szab\'o, M., Nicolis,
  O. and Lakatos, M. (2025) Machine-learning-based probabilistic forecasting of
  solar irradiance in {Chile}.
\newblock \textit{Advances in Statistical Climatology, Meteorology and
  Oceanography}, \textbf{11}, 89--105.

\bibitem[{Baran and Nemoda(2016)}]{bn16}
Baran, S. and Nemoda, D. (2016) Censored and shifted gamma distribution based
  {EMOS} model for probabilistic quantitative precipitation forecasting.
\newblock \textit{Environmetrics}, \textbf{27}, 280--292.

\bibitem[{Benjamini and Hochberg(1995)}]{bh95}
Benjamini, Y. and Hochberg, Y. (1995) Controlling the false discovery rate: A
  practical and powerful approach to multiple testing.
\newblock \textit{Journal of the Royal Statistical Society: Series B
  (Methodological)}, \textbf{57}, 289--300.

\bibitem[{Bi et~al.(2023)Bi, Xie, Zhang, Chen, Gu and Tian}]{BiEtAl2023}
Bi, K., Xie, L., Zhang, H., Chen, X., Gu, X. and Tian, Q. (2023) {Accurate
  medium-range global weather forecasting with 3D neural networks}.
\newblock \textit{Nature}, \textbf{619}, 533--–538.

\bibitem[{Bremnes(2020)}]{b20bqn}
Bremnes, J.~B. (2020) Ensemble postprocessing using quantile function
  regression based on neural networks and {Bernstein} polynomials.
\newblock \textit{Monthly Weather Review}, \textbf{148}, 403--414.

\bibitem[{Bremnes et~al.(2024)Bremnes, Nipen and
  Seierstad}]{bremnes2024evaluation}
Bremnes, J.~B., Nipen, T.~N. and Seierstad, I.~A. (2024) Evaluation of
  forecasts by a global data-driven weather model with and without
  probabilistic post-processing at {Norwegian} stations.
\newblock \textit{Nonlinear Processes in Geophysics}, \textbf{31}, 247--257.

\bibitem[{Br{\"o}cker(2009)}]{b09}
Br{\"o}cker, J. (2009) Reliability, sufficiency, and the decomposition of
  proper scores.
\newblock \textit{Quarterly Journal of the Royal Meteorological Society},
  \textbf{135}, 1512--1519.

\bibitem[{Bröcker(2012)}]{b12}
Bröcker, J. (2012) Evaluating raw ensembles with the continuous ranked
  probability score.
\newblock \textit{Quarterly Journal of the Royal Meteorological Society},
  \textbf{138}, 1611--1617.

\bibitem[{Buizza(2018)}]{b18b}
Buizza, R. (2018) Ensemble forecasting and the need for calibration.
\newblock In \textit{Statistical Postprocessing of Ensemble Forecasts} (eds.
  S.~Vannitsem, D.~S. Wilks and J.~W. Messner), 15--48. Amsterdam: Elsevier.

\bibitem[{B{\"u}lte et~al.(2025)B{\"u}lte, Horat, Quinting and
  Lerch}]{buelte_etal_2024}
B{\"u}lte, C., Horat, N., Quinting, J. and Lerch, S. (2025) Uncertainty
  quantification for data-driven weather models.
\newblock \textit{Artificial Intelligence for the Earth Systems}, in press.

\bibitem[{Cannon(2011)}]{c11hub}
Cannon, A.~J. (2011) Quantile regression neural networks: Implementation in {R}
  and application to precipitation downscaling.
\newblock \textit{Computers and Geosciences}, \textbf{37}, 1277--1284.

\bibitem[{De~Soto et~al.(2006)De~Soto, Klein and Beckman}]{desoto06}
De~Soto, W., Klein, S. and Beckman, W. (2006) Improvement and validation of a
  model for photovoltaic array performance.
\newblock \textit{Solar Energy}, \textbf{80}, 78--88.

\bibitem[{Delle~Monache et~al.(2006)Delle~Monache, Hacker, Zhou, Deng and
  Stull}]{dmhzds06}
Delle~Monache, L., Hacker, J.~P., Zhou, Y., Deng, X. and Stull, R.~B. (2006)
  Probabilistic aspects of meteorological and ozone regional ensemble
  forecasts.
\newblock \textit{Journal of Geophysical Research: Atmospheres}, \textbf{111},
  D24307.

\bibitem[{Demaeyer et~al.(2023)Demaeyer, Bhend, Lerch, Primo, Van~Schaeybroeck,
  Atencia, Ben~Bouall{\`e}gue, Chen, Dabernig, Evans, Faganeli~Pucer, Hooper,
  Horat, Jobst, Mer{\v s}e, Mlakar, M{\"o}ller, Mestre, Taillardat and
  Vannitsem}]{demaeyer_etal_2023_euppbench}
Demaeyer, J., Bhend, J., Lerch, S., Primo, C., Van~Schaeybroeck, B., Atencia,
  A., Ben~Bouall{\`e}gue, Z., Chen, J., Dabernig, M., Evans, G.,
  Faganeli~Pucer, J., Hooper, B., Horat, N., Jobst, D., Mer{\v s}e, J., Mlakar,
  P., M{\"o}ller, A., Mestre, O., Taillardat, M. and Vannitsem, S. (2023) The
  {{EUPPBench}} postprocessing benchmark dataset v1.0.
\newblock \textit{Earth System Science Data}, \textbf{15}, 2635--2653.

\bibitem[{Diebold and Mariano(1995)}]{dm95}
Diebold, F.~X. and Mariano, R.~S. (1995) Comparing predictive accuracy.
\newblock \textit{Journal of Business and Economic Statistics}, \textbf{13},
  253--263.

\bibitem[{Driesse et~al.(2008)Driesse, Jain and Harrison}]{driesse08}
Driesse, A., Jain, P. and Harrison, S. (2008) Beyond the curves: {{Modeling}}
  the electrical efficiency of photovoltaic inverters.
\newblock In \textit{2008 33rd {{IEEE Photovolatic Specialists Conference}}},
  1--6. IEEE.

\bibitem[{Gneiting(2011)}]{g11det}
Gneiting, T. (2011) Making and evaluating point forecasts.
\newblock \textit{Journal of the American Statistical Association},
  \textbf{106}, 746--762.

\bibitem[{Gneiting et~al.(2023{\natexlab{a}})Gneiting, Lerch and
  Schulz}]{gls23}
Gneiting, T., Lerch, S. and Schulz, B. (2023{\natexlab{a}}) Probabilistic solar
  forecasting: Benchmarks, post-processing, verification.
\newblock \textit{Solar Energy}, \textbf{252}, 72--80.

\bibitem[{Gneiting and Raftery(2007)}]{gr07}
Gneiting, T. and Raftery, A.~E. (2007) Strictly proper scoring rules,
  prediction and estimation.
\newblock \textit{Journal of the American Statistical Association},
  \textbf{102}, 359--378.

\bibitem[{Gneiting et~al.(2005)Gneiting, Raftery, Westveld and
  Goldman}]{grwg05}
Gneiting, T., Raftery, A.~E., Westveld, A.~H. and Goldman, T. (2005) Calibrated
  probabilistic forecasting using ensemble model output statistics and minimum
  {CRPS} estimation.
\newblock \textit{Monthly Weather Review}, \textbf{133}, 1098--1118.

\bibitem[{Gneiting and Ranjan(2011)}]{gr11}
Gneiting, T. and Ranjan, R. (2011) Comparing density forecasts using
  thresholdand quantile-weighted scoring rules.
\newblock \textit{Journal of Business and Economic Statistics}, \textbf{29},
  411--422.

\bibitem[{Gneiting et~al.(2023{\natexlab{b}})Gneiting, Wolffram, Resin, Kraus,
  Bracher, Dimitriadis, Hagenmeyer, Jordan, Lerch, Phipps
  et~al.}]{Gneiting2023}
Gneiting, T., Wolffram, D., Resin, J., Kraus, K., Bracher, J., Dimitriadis, T.,
  Hagenmeyer, V., Jordan, A.~I., Lerch, S., Phipps, K. et~al.
  (2023{\natexlab{b}}) Model diagnostics and forecast evaluation for quantiles.
\newblock \textit{Annual Review of Statistics and Its Application},
  \textbf{10}, 597--621.

\bibitem[{Haupt et~al.(2021)Haupt, Chapman, Adams, Kirkwood, Hosking, Robinson,
  Lerch and Subramanian}]{haupt_etal_2021_implementing}
Haupt, S.~E., Chapman, W., Adams, S.~V., Kirkwood, C., Hosking, J.~S.,
  Robinson, N.~H., Lerch, S. and Subramanian, A.~C. (2021) Towards implementing
  artificial intelligence post-processing in weather and climate: Proposed
  actions from the {{Oxford}} 2019 workshop.
\newblock \textit{Philosophical Transactions of the Royal Society A:
  Mathematical, Physical and Engineering Sciences}, \textbf{379}, 20200091.

\bibitem[{Henzi et~al.(2021)Henzi, Ziegel and Gneiting}]{henzi2021isotonic}
Henzi, A., Ziegel, J.~F. and Gneiting, T. (2021) Isotonic distributional
  regression.
\newblock \textit{Journal of the Royal Statistical Society Series B:
  Statistical Methodology}, \textbf{83}, 963--993.

\bibitem[{Hersbach(2000)}]{h00}
Hersbach, H. (2000) Decomposition of the continuous ranked probability score
  for ensemble prediction systems.
\newblock \textit{Weather and Forecasting}, \textbf{15}, 559--570.

\bibitem[{Hong et~al.(2020)Hong, Pinson, Wang, Weron, Yang and
  Zareipour}]{hong2020energy}
Hong, T., Pinson, P., Wang, Y., Weron, R., Yang, D. and Zareipour, H. (2020)
  Energy forecasting: A review and outlook.
\newblock \textit{IEEE Open Access Journal of Power and Energy}, \textbf{7},
  376--388.

\bibitem[{Horat et~al.(2025)Horat, Klerings and Lerch}]{hkl25}
Horat, N., Klerings, S. and Lerch, S. (2025) Improving model chain approaches
  for probabilistic solar energy forecasting through post-processing and
  machine learning.
\newblock \textit{Advances in Atmospheric Sciences}, \textbf{42}, 297--312.

\bibitem[{Jordan et~al.(2019)Jordan, Kr{\"u}ger and Lerch}]{jkl19}
Jordan, A., Kr{\"u}ger, F. and Lerch, S. (2019) Evaluating probabilistic
  forecasts with {scoringRules}.
\newblock \textit{Journal of Statistical Software}, \textbf{90}, 1--37.

\bibitem[{Koenker(2005)}]{koenker2005quantile}
Koenker, R. (2005) \textit{Quantile Regression}.
\newblock Cambridge, UK: Cambridge University Press.

\bibitem[{Kolassa(2020)}]{k20}
Kolassa, S. (2020) Why the “best” point forecast depends on the error or
  accuracy measure.
\newblock \textit{International Journal of Forecasting}, \textbf{36}, 208--211.

\bibitem[{Kr\"uger et~al.(2021)Kr\"uger, Lerch, Thorarinsdottir and
  Gneiting}]{kltg21}
Kr\"uger, F., Lerch, S., Thorarinsdottir, T. and Gneiting, T. (2021) Predictive
  inference based on {Markov Chain Monte Carlo} output.
\newblock \textit{International Statistical Review}, \textbf{89}, 274--301.

\bibitem[{Lam et~al.(2023)Lam, Sanchez-Gonzalez, Willson, Wirnsberger,
  Fortunato, Alet, Ravuri, Ewalds, Eaton-Rosen, Hu, Merose, Hoyer, Holland,
  Vinyals, Stott, Pritzel, Mohamed and Battaglia}]{LamEtAl2023}
Lam, R., Sanchez-Gonzalez, A., Willson, M., Wirnsberger, P., Fortunato, M.,
  Alet, F., Ravuri, S., Ewalds, T., Eaton-Rosen, Z., Hu, W., Merose, A., Hoyer,
  S., Holland, G., Vinyals, O., Stott, J., Pritzel, A., Mohamed, S. and
  Battaglia, P. (2023) Learning skillful medium-range global weather
  forecasting.
\newblock \textit{Science}, \textbf{382}, 1416–1421.

\bibitem[{Lang et~al.(2024{\natexlab{a}})Lang, Alexe, Chantry, Dramsch,
  Pinault, Raoult, Clare, Lessig, Maier-Gerber, Magnusson, Bouallègue,
  Nemesio, Dueben, Brown, Pappenberger and Rabier}]{AIFS}
Lang, S., Alexe, M., Chantry, M., Dramsch, J., Pinault, F., Raoult, B., Clare,
  M. C.~A., Lessig, C., Maier-Gerber, M., Magnusson, L., Bouallègue, Z.~B.,
  Nemesio, A.~P., Dueben, P.~D., Brown, A., Pappenberger, F. and Rabier, F.
  (2024{\natexlab{a}}) {AIFS} -- {ECMWF}'s data-driven forecasting system.
\newblock Preprint, available at \url{https://arxiv.org/abs/2406.01465}.

\bibitem[{Lang et~al.(2024{\natexlab{b}})Lang, Alexe, Clare, Roberts, Adewoyin,
  Bouallègue, Chantry, Dramsch, Dueben, Hahner, Maciel, Prieto-Nemesio,
  O'Brien, Pinault, Polster, Raoult, Tietsche and
  Leutbecher}]{lang2024aifs-crps}
Lang, S., Alexe, M., Clare, M. C.~A., Roberts, C., Adewoyin, R., Bouallègue,
  Z.~B., Chantry, M., Dramsch, J., Dueben, P.~D., Hahner, S., Maciel, P.,
  Prieto-Nemesio, A., O'Brien, C., Pinault, F., Polster, J., Raoult, B.,
  Tietsche, S. and Leutbecher, M. (2024{\natexlab{b}}) {AIFS-CRPS}: Ensemble
  forecasting using a model trained with a loss function based on the
  continuous ranked probability score.
\newblock Preprint, available at \url{https://arxiv.org/abs/2412.15832}.

\bibitem[{Lauret et~al.(2019)Lauret, David and Pinson}]{ldp19}
Lauret, P., David, M. and Pinson, P. (2019) Verification of solar irradiance
  probabilistic forecasts.
\newblock \textit{Solar Energy}, \textbf{194}, 254--271.

\bibitem[{{Le Gal La Salle} et~al.(2020){Le Gal La Salle}, Badosa, David,
  Pinson and Lauret}]{lsb20}
{Le Gal La Salle}, J., Badosa, J., David, M., Pinson, P. and Lauret, P. (2020)
  Added-value of ensemble prediction system on the quality of solar irradiance
  probabilistic forecasts.
\newblock \textit{Renewable Energy}, \textbf{162}, 1321--1339.

\bibitem[{Lefèvre et~al.(2013)Lefèvre, Oumbe, Blanc, Espinar, Gschwind, Qu,
  Wald, Schroedter-Homscheidt, Hoyer-Klick, Arola, Benedetti, Kaiser and
  Morcrette}]{Lefvre2013McClear}
Lefèvre, M., Oumbe, A., Blanc, P., Espinar, B., Gschwind, B., Qu, Z., Wald,
  L., Schroedter-Homscheidt, M., Hoyer-Klick, C., Arola, A., Benedetti, A.,
  Kaiser, J.~W. and Morcrette, J.~J. (2013) {McClear}: A new model estimating
  downwelling solar radiation at ground level in clear-sky conditions.
\newblock \textit{Atmospheric Measurement Techniques}, \textbf{6}, 2403--2418.

\bibitem[{Martin and Ruiz(2001)}]{MartinRuiz2001}
Martin, N. and Ruiz, J. (2001) Calculation of the {PV} modules angular losses
  under field conditions by means of an analytical model.
\newblock \textit{Solar Energy Materials and Solar Cells}, \textbf{70}, 25--38.

\bibitem[{Mattei et~al.(2006)Mattei, Notton, Cristofari, Muselli and
  Poggi}]{Mattei2006}
Mattei, M., Notton, G., Cristofari, C., Muselli, M. and Poggi, P. (2006)
  Calculation of the polycrystalline {PV} module temperature using a simple
  method of energy balance.
\newblock \textit{Renewable Energy}, \textbf{31}, 553--567.

\bibitem[{Mayer(2022)}]{m22impact}
Mayer, M.~J. (2022) Impact of the tilt angle, inverter sizing factor and row
  spacing on the photovoltaic power forecast accuracy.
\newblock \textit{Applied Energy}, \textbf{323}, 119598.

\bibitem[{Mayer and Gr{\'o}f(2020)}]{mg20opt}
Mayer, M.~J. and Gr{\'o}f, G. (2020) Techno-economic optimization of
  grid-connected, ground-mounted photovoltaic power plants by genetic algorithm
  based on a comprehensive mathematical model.
\newblock \textit{Solar Energy}, \textbf{202}, 210--226.

\bibitem[{Mayer and Gr{\'o}f(2021)}]{mg21mc}
--- (2021) Extensive comparison of physical models for photovoltaic power
  forecasting.
\newblock \textit{Applied Energy}, \textbf{283}, 116239.

\bibitem[{Mayer and Yang(2022)}]{my22mce}
Mayer, M.~J. and Yang, D. (2022) Probabilistic photovoltaic power forecasting
  using a calibrated ensemble of model chains.
\newblock \textit{Renewable and Sustainable Energy Reviews}, \textbf{168},
  112821.

\bibitem[{Mayer and Yang(2023)}]{my23cal}
--- (2023) Calibration of deterministic {NWP} forecasts and its impact on
  verification.
\newblock \textit{International Journal of Forecasting}, \textbf{39}, 981--991.

\bibitem[{Mayer and Yang(2024)}]{my24opt}
--- (2024) Optimal place to apply post-processing in the deterministic
  photovoltaic power forecasting workflow.
\newblock \textit{Applied Energy}, \textbf{371}, 123681.

\bibitem[{Messner et~al.(2016)Messner, Mayr and Zeileis}]{mmz16}
Messner, J.~W., Mayr, G.~J. and Zeileis, A. (2016) {Heteroscedastic censored
  and truncated regression with crch}.
\newblock \textit{{The R Journal}}, \textbf{8}, 173--181.

\bibitem[{Messner et~al.(2017)Messner, Mayr and Zeileis}]{mmz17}
--- (2017) Nonhomogeneous boosting for predictor selection in ensemble
  postprocessing.
\newblock \textit{Monthly Weather Review}, \textbf{145}, 137--147.

\bibitem[{Mlakar et~al.(2024)Mlakar, Merše and Faganeli~Pucer}]{Mlakar2024}
Mlakar, P., Merše, J. and Faganeli~Pucer, J. (2024) Ensemble weather forecast
  post‐processing with a flexible probabilistic neural network approach.
\newblock \textit{Quarterly Journal of the Royal Meteorological Society},
  \textbf{150}, 4156–4177.

\bibitem[{Perez et~al.(1990)Perez, Ineichen, Seals, Michalsky and
  Stewart}]{Perez1990trans}
Perez, R., Ineichen, P., Seals, R., Michalsky, J. and Stewart, R. (1990)
  Modeling daylight availability and irradiance components from direct and
  global irradiance.
\newblock \textit{Solar Energy}, \textbf{44}, 271--289.

\bibitem[{Phipps et~al.(2022)Phipps, Lerch, Andersson, Mikut, Hagenmeyer and
  Ludwig}]{PhippsEtAl2022}
Phipps, K., Lerch, S., Andersson, M., Mikut, R., Hagenmeyer, V. and Ludwig, N.
  (2022) Evaluating ensemble post-processing for wind power forecasts.
\newblock \textit{Wind Energy}, \textbf{25}, 1379--1405.

\bibitem[{Politis and Romano(1994)}]{pr94}
Politis, D.~N. and Romano, J.~P. (1994) The stationary bootstrap.
\newblock \textit{Journal of the American Statistical Association},
  \textbf{89}, 1303--1313.

\bibitem[{Price et~al.(2025)Price, Sanchez-Gonzalez, Alet, Andersson, El-Kadi,
  Masters, Ewalds, Stott, Mohamed, Battaglia, Lam and Willson}]{gencast25}
Price, I., Sanchez-Gonzalez, A., Alet, F., Andersson, T.~R., El-Kadi, A.,
  Masters, D., Ewalds, T., Stott, J., Mohamed, S., Battaglia, P., Lam, R. and
  Willson, M. (2025) Probabilistic weather forecasting with machine learning.
\newblock \textit{Nature}, \textbf{637}, 84--90.

\bibitem[{Rasp and Lerch(2018)}]{rl18}
Rasp, S. and Lerch, S. (2018) Neural networks for postprocessing ensemble
  weather forecasts.
\newblock \textit{Monthly Weather Review}, \textbf{146}, 3885--3900.

\bibitem[{Reda and Andreas(2004)}]{ra04spa}
Reda, I. and Andreas, A. (2004) Solar position algorithm for solar radiation
  applications.
\newblock \textit{Solar Energy}, \textbf{76}, 577--589.

\bibitem[{Roberts et~al.(2017)Roberts, Mendiburu~Zevallos and
  Cassula}]{Roberts2017}
Roberts, J.~J., Mendiburu~Zevallos, A.~A. and Cassula, A.~M. (2017) Assessment
  of photovoltaic performance models for system simulation.
\newblock \textit{Renewable and Sustainable Energy Reviews}, \textbf{72},
  1104–1123.

\bibitem[{Scheuerer(2014)}]{sch14}
Scheuerer, M. (2014) Probabilistic quantitative precipitation forecasting using
  ensemble model output statistics.
\newblock \textit{Quarterly Journal of the Royal Meteorological Society},
  \textbf{140}, 1086--1096.

\bibitem[{Schulz et~al.(2021)Schulz, El~Ayari, Lerch and Baran}]{sealb21}
Schulz, B., El~Ayari, M., Lerch, S. and Baran, S. (2021) Post-processing
  numerical weather prediction ensembles for probabilistic solar irradiance
  forecasting.
\newblock \textit{Solar Energy}, \textbf{220}, 1016--1031.

\bibitem[{Schulz et~al.(2022)Schulz, Köhler and Lerch}]{schulz2022aggregating}
Schulz, B., Köhler, L. and Lerch, S. (2022) Aggregating distribution forecasts
  from deep ensembles.
\newblock Preprint, available at \url{https://arxiv.org/abs/2204.02291}.

\bibitem[{Schulz and Lerch(2022)}]{sl22}
Schulz, B. and Lerch, S. (2022) Machine learning methods for postprocessing
  ensemble forecasts of wind gusts: A systematic comparison.
\newblock \textit{Monthly Weather Review}, \textbf{150}, 235--257.

\bibitem[{Song et~al.(2024)Song, Yang, Lerch, Xia, Yagli, Bright, Shen, Liu,
  Liu and Mayer}]{sy24ncqrnn}
Song, M., Yang, D., Lerch, S., Xia, X., Yagli, G.~M., Bright, J.~M., Shen, Y.,
  Liu, B., Liu, X. and Mayer, M.~J. (2024) Non-crossing quantile regression
  neural network as a calibration tool for ensemble weather forecasts.
\newblock \textit{Advances in Atmospheric Sciences}, \textbf{41}, 1417--1437.

\bibitem[{Taillardat et~al.(2016)Taillardat, Mestre, Zamo and
  Naveau}]{tmzn16qrf}
Taillardat, M., Mestre, O., Zamo, M. and Naveau, P. (2016) Calibrated ensemble
  forecasts using quantile regression forests and ensemble model output
  statistics.
\newblock \textit{Monthly Weather Review}, \textbf{144}, 2375--2393.

\bibitem[{Taylor(2000)}]{t00qrnn}
Taylor, J.~W. (2000) A quantile regression neural network approach to
  estimating the conditional density of multiperiod returns.
\newblock \textit{Journal of Forecasting}, \textbf{19}, 299--311.

\bibitem[{Theocharides et~al.(2020)Theocharides, Makrides, Livera, Theristis,
  Kaimakis and Georghiou}]{Theocharides2020}
Theocharides, S., Makrides, G., Livera, A., Theristis, M., Kaimakis, P. and
  Georghiou, G.~E. (2020) Day-ahead photovoltaic power production forecasting
  methodology based on machine learning and statistical post-processing.
\newblock \textit{Applied Energy}, \textbf{268}, 115023.

\bibitem[{Thorarinsdottir and Gneiting(2010)}]{tg10}
Thorarinsdottir, T.~L. and Gneiting, T. (2010) Probabilistic forecasts of wind
  speed: Ensemble model output statistics by using heteroscedastic censored
  regression.
\newblock \textit{Journal of the Royal Statistical Society: Series A
  (Statistics in Society)}, \textbf{173A}, 371--388.

\bibitem[{Van~Schaeybroeck and Vannitsem(2015)}]{van2015ensemble}
Van~Schaeybroeck, B. and Vannitsem, S. (2015) Ensemble post-processing using
  member-by-member approaches: Theoretical aspects.
\newblock \textit{Quarterly Journal of the Royal Meteorological Society},
  \textbf{141}, 807--818.

\bibitem[{Vannitsem et~al.(2021)Vannitsem, Bremnes, Demaeyer, Evans, Flowerdew,
  Hemri, Lerch, Roberts, Theis, Atencia, {Ben Bouallègue}, Bhend, Dabernig,
  Cruz, Hieta, Mestre, Moret, Plenković, Schmeits, Taillardat, den Bergh,
  Schaeybroeck, Whan and Ylhaisi}]{vbd21}
Vannitsem, S., Bremnes, J.~B., Demaeyer, J., Evans, G.~R., Flowerdew, J.,
  Hemri, S., Lerch, S., Roberts, N., Theis, S., Atencia, A., {Ben Bouallègue},
  Z., Bhend, J., Dabernig, M., Cruz, L.~D., Hieta, L., Mestre, O., Moret, L.,
  Plenković, I.~O., Schmeits, M., Taillardat, M., den Bergh, J.~V.,
  Schaeybroeck, B.~V., Whan, K. and Ylhaisi, J. (2021) Statistical
  postprocessing for weather forecasts -- review, challenges and avenues in a
  big data world.
\newblock \textit{Bulletin of the American Meteorological Society},
  \textbf{102}, E681--E699.

\bibitem[{Wang et~al.(2022)Wang, Yang, Hong and Kleissl}]{WangEtAl2022}
Wang, W., Yang, D., Hong, T. and Kleissl, J. (2022) An archived dataset from
  the {ECMWF} ensemble prediction system for probabilistic solar power
  forecasting.
\newblock \textit{Solar Energy}, \textbf{248}, 64--75.

\bibitem[{Wilks(2016)}]{w16}
Wilks, D.~S. (2016) {"The Stippling Shows Statistically Significant Grid
  Points"}: How research results are routinely overstated and overinterpreted,
  and what to do about it.
\newblock \textit{Bulletin of the American Meteorological Society},
  \textbf{97}, 2263--2273.

\bibitem[{Wilks(2019)}]{w19}
--- (2019) \textit{Statistical Methods in the Atmospheric Sciences}.
\newblock Amsterdam: Elsevier, 4th edn.

\bibitem[{Yang(2016)}]{Yang16transrew}
Yang, D. (2016) Solar radiation on inclined surfaces: {{Corrections}} and
  benchmarks.
\newblock \textit{Solar Energy}, \textbf{136}, 288--302.

\bibitem[{Yang(2021)}]{Yang2021septrc}
--- (2021) Temporal-resolution cascade model for separation of 1-min beam and
  diffuse irradiance.
\newblock \textit{Journal of Renewable and Sustainable Energy}, \textbf{13},
  056101.

\bibitem[{Yang(2022)}]{Yang22seprew}
--- (2022) Estimating 1-min beam and diffuse irradiance from the global
  irradiance: {{A}} review and an extensive worldwide comparison of latest
  separation models at 126 stations.
\newblock \textit{Renewable and Sustainable Energy Reviews}, \textbf{159},
  112195.

\bibitem[{Yang et~al.(2024{\natexlab{a}})Yang, Gu, Mayer, Gueymard, Wang,
  Kleissl, Li, Chu and Bright}]{Yang2024sepclust}
Yang, D., Gu, Y., Mayer, M.~J., Gueymard, C.~A., Wang, W., Kleissl, J., Li, M.,
  Chu, Y. and Bright, J.~M. (2024{\natexlab{a}}) Regime-dependent 1-min
  irradiance separation model with climatology clustering.
\newblock \textit{Renewable and Sustainable Energy Reviews}, \textbf{189},
  113992.

\bibitem[{Yang and van~der Meer(2021)}]{ym21pp}
Yang, D. and van~der Meer, D. (2021) Post-processing in solar forecasting: Ten
  overarching thinking tools.
\newblock \textit{Renewable and Sustainable Energy Reviews}, \textbf{140},
  110735.

\bibitem[{Yang et~al.(2022{\natexlab{a}})Yang, Wang, Gueymard, Hong, Kleissl,
  Huang, Perez, Perez, Bright, Xia, van~der Meer and Peters}]{y22ras}
Yang, D., Wang, W., Gueymard, C.~A., Hong, T., Kleissl, J., Huang, J., Perez,
  M.~J., Perez, R., Bright, J.~M., Xia, X., van~der Meer, D. and Peters, I.~M.
  (2022{\natexlab{a}}) A review of solar forecasting, its dependence on
  atmospheric sciences and implications for grid integration: Towards carbon
  neutrality.
\newblock \textit{Renewable and Sustainable Energy Reviews}, \textbf{161},
  112348.

\bibitem[{Yang et~al.(2022{\natexlab{b}})Yang, Wang and Xia}]{ywx22cons}
Yang, D., Wang, W. and Xia, X. (2022{\natexlab{b}}) A concise overview on solar
  resource assessment and forecasting.
\newblock \textit{Advances in Atmospheric Sciences}, \textbf{39}, 1239--1251.

\bibitem[{Yang et~al.(2024{\natexlab{b}})Yang, Xia and Mayer}]{yxm24tut1}
Yang, D., Xia, X. and Mayer, M.~J. (2024{\natexlab{b}}) A tutorial review of
  the solar power curve: Regressions, model chains, and their hybridization and
  probabilistic extensions.
\newblock \textit{Advances in Atmospheric Sciences}, \textbf{41}, 1023--1067.

\bibitem[{Zhong et~al.(2024)Zhong, Chen, Fan, Qian, Liu and Li}]{FuXi2}
Zhong, X., Chen, L., Fan, X., Qian, W., Liu, J. and Li, H. (2024) {FuXi-2.0:
  Advancing machine learning weather forecasting model for practical
  applications}.
\newblock Preprint, available at \url{https://arxiv.org/abs/2409.07188}.

\end{thebibliography}

\appendix

\begin{figure}[!t]
\begin{center}
  \epsfig{file=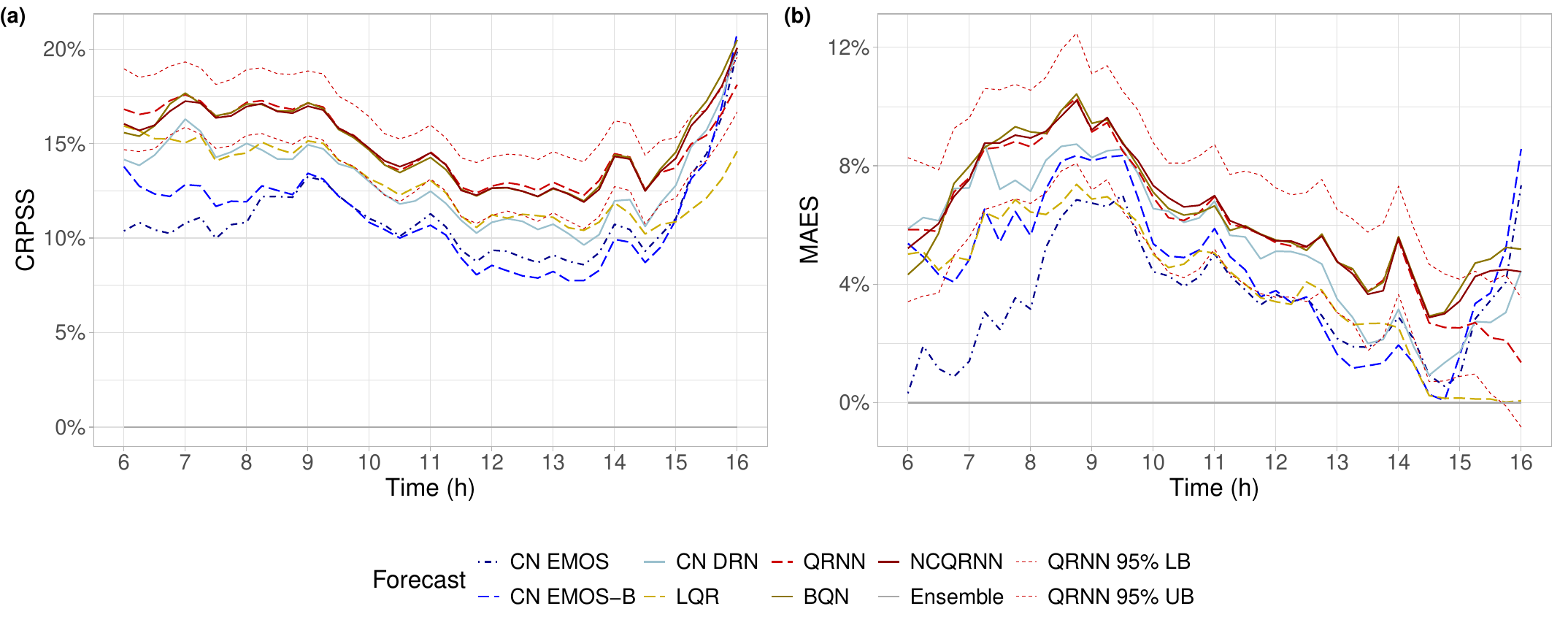, width=\textwidth}
\end{center}
\caption{CRPSS (a) and MAES of the median (b) of post-processed PV power forecasts normalized to the mean daytime power output of the PV plants with respect to the raw ensemble as functions of the observation time (06:00 -- 16:00 UTC) together with 95\,\% confidence intervals for the QRNN method.}
\label{fig:crpss_maesCI}
\end{figure}

\section{Significance of score differences}
\label{secA}

\begin{table}[t]  
  \begin{center}{\footnotesize
    \begin{tabular}{l|ccccccc}
       \hline
      Forecast&Overall&Bodajk&Cegl\'ed&Fels\H ozsolca
\\ \hline
CN EMOS&22.41$\pm$0.23\%&23.13$\pm$0.63\%&22.37$\pm$0.53\%&24.77$\pm$0.71\% \\
CN EMOS–B&22.44$\pm$0.23\%&23.10$\pm$0.64\%&22.26$\pm$0.53\%&24.84$\pm$0.72\% \\
CN DRN&21.92$\pm$0.23\%&22.36$\pm$0.62\%&21.79$\pm$0.55\%&24.15$\pm$0.74\% \\ \hline
LQR&21.90$\pm$0.23\%&22.57$\pm$0.61\%&21.64$\pm$0.53\%&24.21$\pm$0.69\%\\
QRNN&21.42$\pm$0.23\%&21.99$\pm$0.61\%&21.23$\pm$0.54\%&23.49$\pm$0.72\% \\
BQN&21.44$\pm$0.23\%&22.00$\pm$0.61\%&21.22$\pm$0.55\%&23.46$\pm$0.73\% \\
NCQRNN&21.44$\pm$0.23\%&22.03$\pm$0.61\%&21.21$\pm$0.54\%&23.59$\pm$0.72\%\\ \hline
Ensemble&25.08$\pm$0.26\%&26.09$\pm$0.71\%&24.83$\pm$0.61\%&27.58$\pm$0.78\% \\ \hline \hline
&Fert\H osz\'eplak&Magyarsarl\'os&Paks&\'Ujk\'{\i}gy\'os
\\ \hline
CN EMOS&23.62$\pm$0.69\%&20.47$\pm$0.54\%&20.96$\pm$0.59\%&21.63$\pm$0.57\% \\
CN EMOS–B&23.81$\pm$0.69\%&20.41$\pm$0.54\%&20.94$\pm$0.58\%&21.80$\pm$0.57\%\\
CN DRN&23.14$\pm$0.69\%&20.19$\pm$0.54\%&20.52$\pm$0.59\%&21.35$\pm$0.59\% \\ \hline
LQR&23.16$\pm$0.66\%&20.28$\pm$0.53\%&20.35$\pm$0.58\%&21.15$\pm$0.56\%\\
QRNN&22.57$\pm$0.68\%&19.95$\pm$0.54\%&20.04$\pm$0.57\%&20.71$\pm$0.57\% \\
BQN&22.63$\pm$0.69\%&20.00$\pm$0.53\%&20.02$\pm$0.58\%&20.81$\pm$0.58\% \\
NCQRNN&22.67$\pm$0.68\%&19.85$\pm$0.54\%&20.04$\pm$0.58\%&20.75$\pm$0.58\% \\ \hline
Ensemble&25.93$\pm$0.79\%&24.66$\pm$0.61\%&23.08$\pm$0.65\%&23.45$\pm$0.64\%\\ \hline
       \end{tabular}
       }
    \end{center}
    \caption{Mean CRPS of post-processed and raw PV power forecasts normalized to the
mean daytime power output of the PV plants for the 06:00 -- 16:00 UTC time period together with 95\,\% confidence intervals.}
    \label{tab:crpsCI}
\end{table}

To address the significance of differences between the various forecasts in terms of the mean CRPS and MAE, we consider two different approaches. On the one hand, we complement the mean scores and some of the skill scores with Gaussian 95\,\% confidence intervals using standard deviations based on 2000 stationary block bootstrap samples with random block lengths drawn from a geometric distribution \citep{pr94}. On the other hand, for each location and observation time, we perform pairwise Diebold-Mariano \citep[DM;][]{dm95} tests for equal predictive performance and report the proportion of cases where the difference in mean CRPS and MAE is significant at a 5\,\% level. Following the suggestions of \citet{w16}, to control the false discovery rate in simultaneous testing, we apply the Benjamini-Hochberg algorithm \citep{bh95}. To avoid the distortion resulting from very low observed and forecasted PV output values, the following analysis restricts the time interval of observations to the hours of peak PV power production between 06:00 and 16:00 UTC.

According to Figure \ref{fig:crpss_maesCI}a, the parametric approaches and the simple LQR are significantly behind the QRNN in terms of the CRPSS for almost all considered observation times, and the same applies for the other two advanced nonparametric methods (BQN and NCQRNN, not shown). Considering the MAES of the median (Figure \ref{fig:crpss_maesCI}b), the situation slightly changes, as the skill scores of the CN DRN approach are mainly between the lower and upper confidence bounds for the MAES of the QRNN (and the BQN and NCQRNN as well, not shown). Finally, even the CRPSS values of the worst performing CN EMOS approach are significantly positive during the whole observation period (not shown), the minimal value of the 95\,\% lower bound is 7.27\,\%, and the MAES of this parametric method is not signficantly positive at a 5\,\% level only at 06:00, 06:30, 06:45 and 14:30, 14:45, 15:00 UTC (not shown).

\begin{table}[t]  
  \begin{center}{\footnotesize
    \begin{tabular}{l|ccccccc}
       \hline
      Forecast&Overall&Bodajk&Cegl\'ed&Fels\H ozsolca
\\ \hline
CN EMOS&31.31$\pm$0.33\%&32.32$\pm$0.90\%&31.58$\pm$0.78\%&34.51$\pm$1.00\% \\
CN EMOS–B&31.03$\pm$0.32\%&31.94$\pm$0.90\%&31.06$\pm$0.76\%&34.34$\pm$0.99\% \\
CN DRN&30.69$\pm$0.33\%&31.49$\pm$0.89\%&30.77$\pm$0.79\%&33.81$\pm$1.03\% \\ \hline
LQR&31.11$\pm$0.33\%&32.10$\pm$0.90\%&30.86$\pm$0.80\%&34.32$\pm$1.01\%\\
QRNN&30.46$\pm$0.34\%&31.38$\pm$0.88\%&30.25$\pm$0.81\%&33.23$\pm$1.06\% \\
BQN&30.40$\pm$0.34\%&31.26$\pm$0.88\%&30.17$\pm$0.82\%&33.04$\pm$1.07\% \\
NCQRNN&30.40$\pm$0.34\%&31.32$\pm$0.89\%&30.17$\pm$0.81\%&33.28$\pm$1.05\%\\ \hline
Ensemble&32.53$\pm$0.34\%&33.89$\pm$0.92\%&32.20$\pm$0.83\%&35.57$\pm$1.02\% \\ \hline \hline
&Fert\H osz\'eplak&Magyarsarl\'os&Paks&\'Ujk\'{\i}gy\'os
\\ \hline
CN EMOS&33.12$\pm$0.95\%&28.12$\pm$0.78\%&29.31$\pm$0.85\%&30.29$\pm$0.84\% \\
CN EMOS–B&32.98$\pm$0.95\%&27.85$\pm$0.76\%&28.95$\pm$0.81\%&30.20$\pm$0.83\%\\
CN DRN&32.47$\pm$0.96\%&27.89$\pm$0.78\%&28.75$\pm$0.85\%&29.78$\pm$0.84\% \\ \hline
LQR&32.97$\pm$0.99\%&29.11$\pm$0.78\%&28.77$\pm$0.85\%&29.77$\pm$0.86\%\\
QRNN&32.15$\pm$1.00\%&28.50$\pm$0.80\%&28.47$\pm$0.84\%&29.36$\pm$0.87\% \\
BQN&32.09$\pm$0.99\%&28.65$\pm$0.80\%&28.37$\pm$0.85\%&29.33$\pm$0.86\% \\
NCQRNN&32.11$\pm$0.99\%&28.36$\pm$0.80\%&28.41$\pm$0.85\%&29.26$\pm$0.86\% \\ \hline
Ensemble&33.56$\pm$1.01\%&32.05$\pm$0.81\%&30.10$\pm$0.88\%&30.43$\pm$0.85\%\\ \hline
       \end{tabular}
       }
    \end{center}
    \caption{MAE of the median of post-processed and raw PV power forecasts normalized to the
mean daytime power output of the PV plants for the 06:00 -- 16:00 UTC time period together with 95\,\% confidence intervals.}
    \label{tab:maeCI}
\end{table}

Furthermore, Tables \ref{tab:crpsCI} and \ref{tab:maeCI} confirm that compared to the raw ensemble, any form of post-processing significantly and consistently improves the mean CRPS of the probabilistic predictions and the MAE of the median forecasts. They also verify that where Tables \ref{tab:crps} and \ref{tab:locations} show substantial differences in terms of the mean CRPS and MAE between the QRNN, BQN, and NCQRNN approaches and the other four post-processing methods, these differences are significant at a 5\,\% level.

\begin{figure}[t]
\begin{center}
\epsfig{file=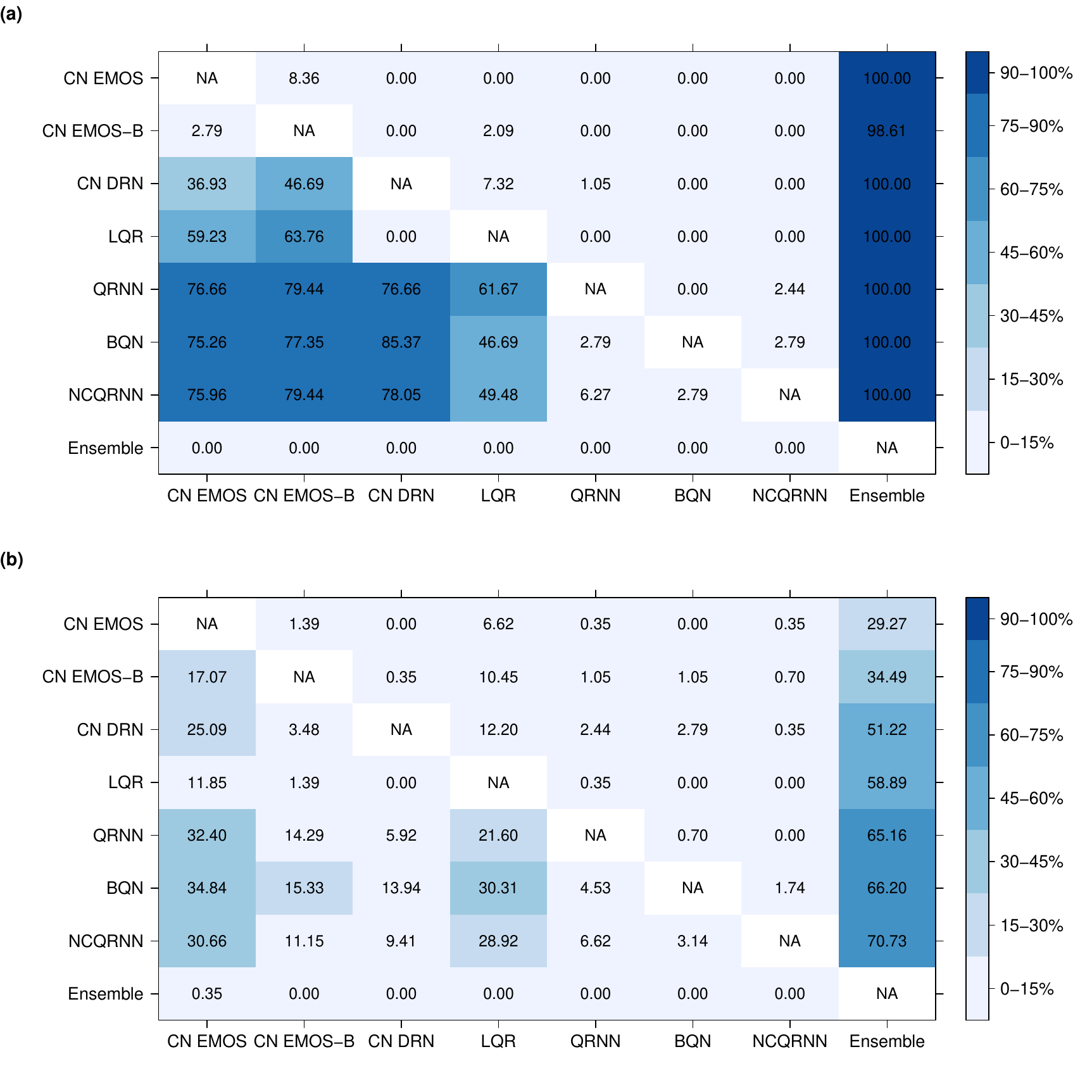, width=\textwidth}
\end{center}
\caption{Proportion of cases where the null hypothesis of equal predictive performance in terms of the mean CRPS (a) and MAE (b) of the corresponding one-sided DM test is rejected at a 5\,\% level of significance in favor of the forecast in the row when compared with the forecast
in the column. }
\label{fig:dmtest}
\end{figure}

Finally, Figure \ref{fig:dmtest} approaches the question of significance in the score differences from another angle. Each entry summarizes the results of 287 parallel pairwise one-sided DM tests (7 locations, 41 observation times) by reporting the proportion of cases where the difference in predictive performance of the compared forecasts is significant at a 5\,\% level. The results of these DM tests are completely in line with the confidence interval-based finding: the differences between QRNN, BQN, and NCQRNN are minor, the raw forecast is significantly behind the post-processed ones in more than 98\,\% of the cases, and the differences between the various post-processing approaches in terms of the MAE of the median are less pronounced than in terms of the mean CRPS.  

\end{document}